\documentclass[preprint,notoc]{JHEP3}
\usepackage{epsfig}

\def\beas{\begin{eqnarray*}}  
\def\eeas{\end{eqnarray*}}

\def\eslt{E_T^{\rm miss}}

\def\to{\rightarrow}

\def\bi{\begin{itemize}}
\def\ei{\end{itemize}}
\def\te{\tilde e}

\def\th{\tilde h}

\def\tu{\tilde u}

\def\tb{\tilde b}

\def\tst{\tilde t}
\def\ttau{\tilde{\tau}}

\def\tg{\tilde g}
\def\tnu{\tilde\nu}

\def\tw{\widetilde W}
\def\tz{\widetilde Z}
\def\delew{\Delta_{EW}}
\def\alt{\lesssim}

\def\be{\begin{equation}}  
\def\ee{\end{equation}}  
\def\bea{\begin{eqnarray}}  
\def\eea{\end{eqnarray}}

\title{Physics at a Higgsino Factory
}
\author{Howard Baer$^a$, Vernon Barger$^b$, Dan Mickelson$^a$, 
Azar Mustafayev$^c$ and Xerxes Tata$^c$\\
$^a$Dept.\ of Physics and Astronomy, University of Oklahoma, Norman, OK 73019, USA\\
$^b$Dept. of Physics, University of Wisconsin, Madison, WI 53706, USA\\
$^c$Dept.\ of Physics and Astronomy, University of Hawaii, Honolulu, HI 96822, USA\\
E-mail: \email{baer@nhn.ou.edu}, \email{barger@pheno.wisc.edu}, 
\email{mickelso@nhn.ou.edu}, \email{azar@phys.hawaii.edu}, \email{tata@phys.hawaii.edu}
}

\preprint{\vbox{UH-511-1234-14}}

\abstract{Naturalness arguments applied to supersymmetric theories imply
a spectrum containing four light higgsinos $\tz_{1,2}$ and $\tw_1^\pm$
with masses $\sim 100-300$~GeV (the closer to $M_Z$, the more natural).
The compressed mass spectrum and associated low energy release from $\tw_1$
and $\tz_2$ three-body decay makes higgsinos difficult to detect at
LHC14, while the other sparticles might be heavy, and possibly even beyond
LHC14 reach.  In contrast, the International Linear $e^+e^-$
Collider (ILC) with $\sqrt{s}>2m(higgsino)$ would be a {\it higgsino
factory} in addition to a Higgs boson factory and would serve as a
discovery machine for natural SUSY!  In this case, both chargino and
neutralino production 
occur at comparable rates, and lead to observable signals above SM
backgrounds. We examine two benchmark cases, one just beyond the LHC8 reach 
with $\tw_1(\tz_2)-\tz_1$ mass gap of 15 (21)~GeV, 
and a second more difficult case beyond even the
LHC14 reach, where the mass gap is just 10~GeV, close to its minimum in
models with no worse than 3\% fine-tuning.
The signal is characterized by low visible energy events together with $\eslt$ 
in the one or two jets $+1\ell$ channel from chargino production, 
and in the opposite sign, same-flavour, acoplanar dilepton
channel from $\tz_1\tz_2$ production. For both cases, we find that the
signal is observable above backgrounds from the usual $2\to 2$ SM events
and from $\gamma\gamma$ collisions with just a few fb$^{-1}$ of
integrated luminosity. We also show that with an integrated luminosity
of 100~fb$^{-1}$, it should be possible to extract $\tw_1$ and $\tz_1$
masses at 2-3\% level from chargino events if the mass gap is $\ge
15$~GeV, and neutralino masses at the sub-percent level from neutralino
events. The latter should also allow a determination of
$m_{\tz_2}-m_{\tz_1}$ at the 200~MeV level. These measurements would
point to higgsinos as the origin of new physics and strongly suggest a link to a
natural origin for $W$, $Z$ and $h$ masses. 

} 

\begin{document}

\section{Introduction: naturalness and light higgsinos}
\label{sec:intro}

Now that a (supposedly fundamental) scalar boson has been discovered
with mass $m_h=125.5\pm 0.5$~GeV at LHC\cite{atlas_h,cms_h}, the puzzle
is: why is it so light?  It is well known that radiative corrections
pull elementary scalar masses up to 
the mass of the heaviest particle to which the Higgs boson couples; 
 {\it e.g.} to the GUT scale in the context of Grand Unified theories. 
Theories with supersymmetry softly broken at the weak scale have much
better ultra-violet behaviour and so allow
for the  co-existence of the weak energy scale
with  $M_{\rm GUT}$.  But so far, there
is no sign of supersymmetric matter at LHC8\cite{atlas_susy,cms_susy}.
This latter issue has given rise to a supposed crisis in supersymmetric
naturalness, and has lead some authors to question whether the simple
picture of weak scale supersymmetry  addresses the issue of naturalness in
a satisfactory manner \cite{Shifman:2012na,craig}.

Such a discussion depends on the measure of supersymmetric naturalness 
which is adopted and how it is applied.  Here, we will adopt the quantity
$\Delta_{EW}$~\cite{ltr,DEWsug,rns,snowmass1,comp,am_xt,dew} that leads
to a model-independent bound on the magnitude of the superpotential
parameter $|\mu|$ which is the higgsino mass in most models.
\footnote{ The authors of Ref.~\cite{comp,dew} argue that the commonly
adopted ``large log'' measure which favors light third generation
squarks $m_{\tst_{1,2},\tb_1}\alt 500$~GeV neglects a variety of
non-independent contributions, leading to over-estimates of fine-tuning
by orders of magnitude.  Once dependent terms are collected, this
measure reduces to $\Delta_{EW}$.  They also argue that the traditional
$\Delta_{BG}$ measure of fractional change in $M_Z^2$ against fractional
change in model parameters is highly model-dependent for multi-parameter
SUSY theories but reduces to $\Delta_{EW}$ when all soft parameters are
correlated and expressed in terms of a single parameter such as the
gravitino mass $m_{3/2}$. See Ref.~\cite{comp,dew} for a detailed
discussion. In contrast, in Ref.~\cite{am_xt} $\delew^{-1}$ is regarded
only as an {\em upper bound} on the fine-tuning in any model with a
specified spectrum. For the purposes of this paper, it does not matter
which view one adopts. It is only important that a value $|\mu|$ close
to $M_Z$ is necessary for low fine-tuning.} To construct $\delew$, we
begin with the scalar potential minimization condition including
radiative corrections:
\be 
\frac{M_Z^2}{2} = \frac{m_{H_d}^2+\Sigma_d^d - (m_{H_u}^2+\Sigma_u^u) \tan^2\beta}{\tan^2\beta -1} -\mu^2 ,
\label{eq:mZsSig}
\ee 
and require that the observed value of $M_Z^2$ is obtained
without  large cancellations between terms on the right-hand-side.
Since all entries in Eq.~(\ref{eq:mZsSig}) are defined at the weak scale, 
the {\it  electroweak fine-tuning measure} is defined by,
\be 
\Delta_{EW} \equiv max_i \left|C_i\right|/(M_Z^2/2)\;, 
\ee 
where $C_{H_d}=m_{H_d}^2/(\tan^2\beta -1)$,
$C_{H_u}=-m_{H_u}^2\tan^2\beta /(\tan^2\beta -1)$ and $C_\mu =-\mu^2$.
Also, $C_{\Sigma_u^u(k)} =-\Sigma_u^u(k)\tan^2\beta /(\tan^2\beta -1)$
and $C_{\Sigma_d^d(k)}=\Sigma_d^d(k)/(\tan^2\beta -1)$, where $k$ labels
the various loop contributions included in Eq.~(\ref{eq:mZsSig}).  A low
value of $\Delta_{EW}\alt 10-30$ means that all contributions to $M_Z$
(or equivalently, to $m_h$) are comparable to $M_Z$ (or $m_h$), {\it
i.e.} no large cancellation between terms is required to generate the
mass scale $M_W,\ M_Z,\ m_h\sim 100$~GeV.

In order to generate low $\Delta_{EW}\sim 10-30$,\footnote{The
importance of low $|\mu |\sim M_Z$ was emphasized in Ref.~\cite{ccn}.
The first of Ref's.~\cite{fp} also remarks that there be no large cancellation between
$m_{H_u}^2$ and $\mu^2$.
Ref's~\cite{golden,perel,martin,nevzorov,guido} also adopt weak scale
fine-tuning for their discussion.  Ref.~\cite{ltr} introduces $\Delta_{EW}$
including finite radiative corrections, and notes that large $A_t$ suppresses
radiative corrections while lifting the value of $m_h$.} 
from Eq.~(\ref{eq:mZsSig}) we see that:
\bi
\item $|\mu |\sim 100-300$~GeV, the lower the better
\item $m_{H_u}^2$ is driven radiatively to small (usually negative) values, and
\item $\Sigma_u^u (\tst_{1,2})\alt 100-300$~GeV.  
\ei 
Here, the lower end of the range of $|\mu |$ comes from the LEP2 bound on the
chargino, while the upper end comes from requiring $\delew \lesssim
3$\%.  
The last requirement can be met when $m_{\tst_1}\sim 1-2$~TeV with
$m_{\tst_2}\sim 2-4$~TeV with large mixing due to a large value of
$A_t$.  A large trilinear SUSY breaking term $A_t$ suppresses both the
$\tst_1$ and $\tst_2$ radiative corrections whilst lifting $m_h$ into
the 125~GeV regime\cite{ltr,rns}.  The required low values of
$\Sigma_u^u(\tst_{1,2})$ mean the top-squarks cannot be too heavy: in
this case, much above a few TeV, but nevertheless likely too heavy to be revealed in
LHC searches. This is in sharp contrast to  
expectations from previous analyses of natural SUSY models, 
and in accord with 
1. a light Higgs boson mass $m_h\sim 125$~GeV, 
2. suppression of anomalous contributions to the $b\to s\gamma$ branching fraction (which is in near
agreement with the SM value) and 
3. lack of a signal from light stops at LHC8.  
The gluino mass, which contributes radiatively to
$m_{\tst_{1,2}}$, is also then bounded from above~\cite{brust}.  
For $\Delta_{EW}\alt 10$ (30), then $m_{\tg}\alt 2$~TeV (4~TeV)\cite{rns}.

Thus, spectra from low $\Delta_{EW}$ models are characterized by:
\bi
\item four light higgsinos $\tw_1^\pm$, $\tz_1$ and $\tz_2$ with mass $\sim \mu\sim 100-300$~GeV,
\item well-mixed top and bottom squarks in the few TeV regime,
\item $m_{\tg}\alt 2-4$~TeV and
\item first/second generation squarks and sleptons in the $5-30$~TeV
regime\footnote{Large first/second generation squark and slepton masses
can again lead to violations of naturalness due to EW $D$-term
contributions unless one of several conditions leading to degeneracy
patterns within GUT multiplets {\it etc.} is respected\cite{deg}.},
consistent with at least a partial decoupling solution to the SUSY
flavor, $CP$, gravitino and $p$-decay problems\cite{dine}.
\ei
Models with such spectra have been described as {\it radiatively-driven
natural supersymmetry}, or RNS, since the value of $m_{H_u}^2$ is
radiatively driven to values close to $M_Z^2$.  RNS spectra can be
realized in the 2-extra-parameter non-universal Higgs models (NUHM2),
but not in more constrained models such as mSUGRA/CMSSM.  For the case
of mSUGRA, while $\mu$ can become low in the HB/FP region, the rather
heavy top squarks $m_{\tst_{1,2}}\sim 7-15$~TeV produce large
$\Sigma_u^u(\tst_{1,2})$ leading again to substantial fine-tuning\cite{DEWsug}.

Light higgsinos can be produced at large rates at LHC8 and LHC14
\cite{lhc}.  However, the compressed higgsino spectrum leads to only small
visible energy release from $\tw_1\to\tz_1 f\bar{f}'$ and $\tz_2\to
\tz_1 f\bar{f}$ decays (where $f$ denotes SM fermions).
LHC14 should probe gluinos with mass up to $m_{\tg}\sim 2$~TeV, assuming an
integrated luminosity of $\sim 1000$~fb$^{-1}$. This means that LHC14 probes
about half of the gluino mass range allowed by RNS\cite{lhc}.  A
distinctive wino pair production signal from $pp\to\tw_2^\pm\tz_4X$
followed by $\tw_2\to W\tz_{1,2}$ and $\tz_4\to W^\pm\tw_1^\mp$ leads to
a novel same-sign diboson signature which, assuming gaugino mass
unification, gives a somewhat better reach than the gluino pair
production channel, and is distinctive in SUSY models with light
higgsinos \cite{SSdB}.  A possible signal from monojets recoiling against
nearly invisible higgsino pairs can occur at the 1\% level above QCD
background for the high luminosity LHC14 \cite{chan,mono,sz}; this does
not appear to us to be enough to claim a discovery potential, though a study
\cite{kribs} has suggested discovery might be possible by examining the
soft-lepton daughters of $\tw_1$ and $\tz_2$.

In this paper we examine the detectability of the light higgsinos of RNS
at the International Linear Collider (ILC), a proposed $e^+e^-$
collider\cite{tdr,snowmass} designed to operate at $\sqrt{s}\sim 0.25-1$
TeV, with an added capability of electron beam polarization.  While such
a machine is often touted as a Higgs boson factory due to the capacity
to study the reaction $e^+e^-\to Zh$, for the case of models with light
higgsinos that are required for naturalness,\footnote{We assume that the
superpotential $\mu$-term that enters via the scalar Higgs potential
also makes the dominant contribution to the higgsino mass. This is the
case in all models that we are aware of. However, in models without any
additional singlet that couples to higgsinos, a SUSY-breaking higgsino
mass term would be soft \cite{girardello}, and the connection between
the higgsino mass and the scalar Higgs potential is lost. See also
Ref.~\cite{brust}. }  the ILC would also become a {\it higgsino factory}
and a {\it SUSY discovery machine}\cite{bbh}!

We remark here that early pioneering studies were performed by the JLC
group on mixed higgsino-wino type of chargino pair production where mass
gaps were around 50~GeV\cite{jlc}.  Additional studies incorporating
cascade decays were performed in Ref.~\cite{bmt}, and in
Ref. \cite{tadas1,tadas2} chargino pair production in the hyperbolic
branch/focus point region \cite{ccn,fp} was examined also with $\sim
40$~GeV mass gaps.  Very recent studies include those in
Ref.~\cite{Han:2013kza}.

Recently, studies of higgsino pair production with mass gaps of order 1
GeV have been performed\cite{Berggren:2013vfa}.  In these studies, use
is made of initial state photon radiation and exclusive one-or-two
particle hadronic decays of the charginos which have large branching
fractions because the $Q$-value is limited at the GeV-level.  These
studies were relevant for string-motivated high-scale gauge-mediation
models where the very large gaugino masses lead to uncomfortably large
values of $\Delta_{EW}\sim 275$\cite{ilcbm}.  In the current paper, we
examine the case of models with $\Delta_{EW}\sim 10-30$ where mass gaps
of 10-20~GeV are typical, and for which the techniques of
Ref.~\cite{Berggren:2013vfa} are not needed.

In Sec.~\ref{sec:bm}, we present two RNS benchmark models labeled as
ILC1 and ILC2.  In Sec.~\ref{sec:prod}, we discuss sparticle production
and decay for the two RNS benchmark models at a higgsino factory.  In
Sec.~\ref{sec:evgen}, we present some details of our signal and
background event generation calculations.  In Sec.~\ref{sec:det}, we
discuss how ILC acts as a ``natural SUSY'' discovery machine for light
higgsino pair production and show how it can make precision measurements
of the associated sparticle masses and mass gaps.  We conclude in
Sec.~\ref{sec:conclude}.

\section{Two RNS benchmark points}
\label{sec:bm}

We select two benchmark points for our study of light higgsinos in RNS
models.  To generate spectra for models with low $\Delta_{EW}$, we use
the Isasugra spectrum generator\cite{isasugra} from Isajet~7.84\cite{isajet}.  
Isajet versions $\ge 7.83$ compute a value of
$\Delta_{EW}$ for each SUSY spectrum. For both benchmark models, we
input parameters from the NUHM2 model\cite{nuhm2}, as listed in Table
\ref{tab:bm}.  Point ILC1 is similar to benchmark RNS2 of
Ref.~\cite{ltr}, except a lower $\mu =115$~GeV value has been selected
to yield a spectrum of light higgsinos which would already be accessible
at ILC250 (ILC with $\sqrt{s}=250$~GeV). The mass gaps for ILC1 are
$m_{\tw_1}-m_{\tz_1}=14.6$~GeV and $m_{\tz_2}-m_{\tz_1}=21.3$~GeV. While
safe from LHC8 bounds, gluino and also wino production will lead to
observable signals at LHC14\cite{lhc}. 

We also examine a much more challenging case of benchmark ILC2 which
will likely be beyond the reach of LHC14. This point is chosen from the
RNS model-line with $\mu=150$~GeV introduced in Ref.~\cite{lhc}, with
$m_{1/2}$ adjusted to obtain as small a mass gap as possible, consistent
with $\delew^{-1}$ of no more than 3\%. For this challenging case, the
mass gaps are rather small, with $m_{\tw_1}-m_{\tz_1}=10.2$~GeV and
$m_{\tz_2}-m_{\tz_1}=9.7$~GeV.  This point is not accessible to ILC250,
so that we examine the feasibility of detection at a centre-of-mass
energy just below the top pair threshold.

We stress that, within the RNS framework, $m_{\tz_2}-m_{\tz_1} =10$~GeV
is close to the minimum of the mass gap if we require that
$\delew^{-1}>3$\%. This can be seen from Fig.~\ref{fig:gap} where we
show the neutralino mass gap in the $m_{1/2}-\mu$ plane, with the other
NUHM2 parameters fixed at the same values as for the ILC2 case.  We see
from the figure that the mass gap ranges from about 10~GeV (for large
$m_{1/2}$, where the $\Delta_{EW}$ contours become vertical because the
top squarks become too heavy) to over 100~GeV in the region where the
gaugino and higgsino states are strongly mixed. 
\FIGURE[tbh]{
\includegraphics[width=12cm,clip]{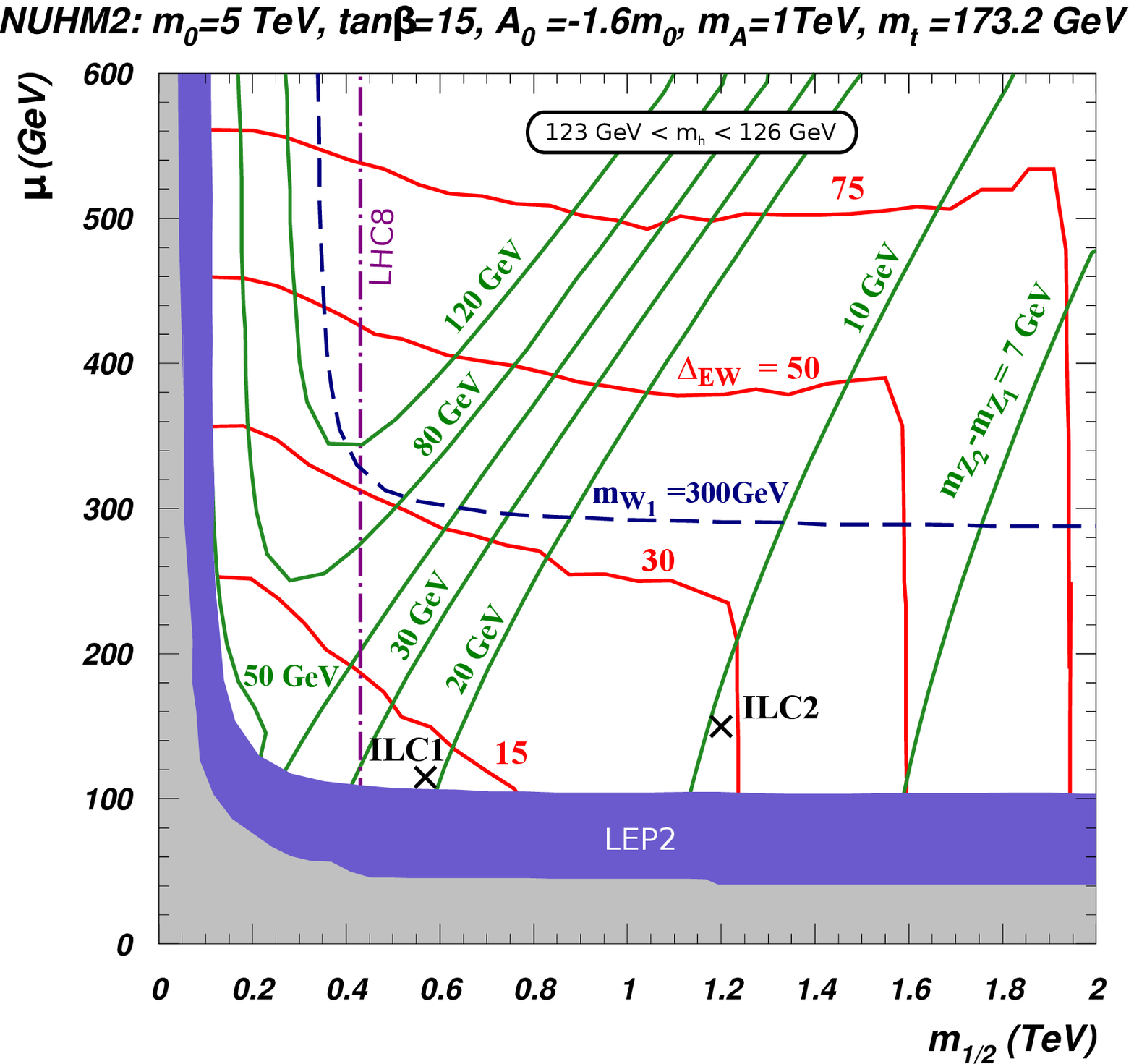}
\caption{Contours of the mass gap (green curves) $m_{\tz_2}-m_{\tz_1}$
  in the $m_{1/2}-\mu$ mass plane of the NUHM2 model for $m_0=5$~TeV,
  $A_0=-1.6m_0$, $\tan\beta=15$ and $m_A=1$~TeV. The red curves show
  contours of $\Delta_{EW}$. The blue (gray) shaded regions are 
  excluded by the absence of a chargino signal at LEP2 (LEP1). The
  region to the left of the dot-dashed line is excluded by the LHC8 limit
  $m_{\tg} > 1.2$~TeV, obtained assuming squarks are very heavy. The
  dashed line is where $m_{\tw_1}=300$~GeV. 
  The crosses denote $\mu$ and $m_{1/2}$ values for ILC1 and ILC2 benchmark points. 
  Note that the other parameters for ILC1 differ from those in the figure, 
  but the mass difference is insensitive to these.}
\label{fig:gap}}
For these large mass
gaps, LHC experiments should be awash in clean multilepton signals from
electroweak-ino production, including signals from $\tw_2$ and $\tz_4$
events.\footnote{Within the RNS framework where gaugino mass parameters
are assumed to be unified, this region is excluded by the LHC gluino
search. Electroweak-ino production can nonetheless serve to probe more
general models.}  Our concern here, of course, is the difficult region
with a small mass gaps $m_{\tz_2}-m_{\tz_1}$ and $m_{\tw_1}-m_{\tz_1}$
where there may well be no detectable signals even at LHC14.  To the
extent that the difficulty of extracting ILC SUSY signals (without using
kinematic properties particular to exclusive chargino decays
\cite{Berggren:2013vfa}) increases with decreasing mass gap, the ILC2
point represents nearly the most challenging case that we may encounter
in our examination of linear colliders as a definitive probe of
naturalness.

We list at the bottom of Table~\ref{tab:bm} the neutralino relic density,
some $B$-decay branching fractions
and WIMP detection rates along with the value of $\Delta_{EW}$.
WIMP detection sensitivities should be multiplied by a factor $\xi\equiv
\Omega_{\tz_1}h^2/0.12$ since the higgsino-like WIMPs could make up only
a fraction of the local DM density\cite{bbm}, whilst {\it e.g.} axions
might make up the remainder\cite{dfsz}.
\begin{table}\centering
\begin{tabular}{lcc}
\hline
parameter & ILC1 & ILC2 \\
\hline
$m_0$      & 7025.0 & 5000  \\
$m_{1/2}$  &  568.3 & 1200  \\
$A_0$      & -11426.6 & -8000  \\
$\tan\beta$&  10 & 15  \\
$\mu$      & 115 & 150  \\
$m_A$      &  1000 & 1000  \\
\hline
$m_{\tg}$   & 1563.5 & 2832.6   \\
$m_{\tu_L}$ &  7021.3 & 5440.4  \\
$m_{\tu_R}$ &  7254.2 & 5565.6  \\
$m_{\te_R}$ &  6758.6 & 4817.1  \\
$m_{\tst_1}$&  1893.3 & 1774.3  \\
$m_{\tst_2}$&  4919.4 & 3877.9  \\
$m_{\tb_1}$ &  4959.2 & 3902.8  \\
$m_{\tb_2}$ &  6893.3 & 5204.5  \\
$m_{\ttau_1}$ & 6656.6 & 4652.3  \\
$m_{\ttau_2}$ & 7103.1 & 5072.5  \\
$m_{\tnu_{\tau}}$ & 7114.0 & 5078.7 \\
$m_{\tw_2}$ & 513.0 & 1017.5  \\
$m_{\tw_1}$ & 117.3 & 158.3  \\
$m_{\tz_4}$ & 524.2 & 1031.1  \\ 
$m_{\tz_3}$ & 267.0 & 538.7   \\ 
$m_{\tz_2}$ & 124.0 & 157.8  \\ 
$m_{\tz_1}$ & 102.7 & 148.1  \\ 
$m_h$      & 125.3 & 125.4  \\ 
\hline
$\Omega_{\tz_1}^{std}h^2$ & 0.009 & 0.007  \\
$BF(b\to s\gamma)\times 10^4$ & 3.3 & $3.3$  \\
$BF(B_s\to \mu^+\mu^-)\times 10^9$ & 3.8 & $3.9$  \\
$\sigma^{SI}(\tz_1 p)$ (pb) & $1.3\times 10^{-8}$ & $2.9\times 10^{-9}$ \\
$\Delta_{EW}$ & 13.9 & 28.5 \\
\hline
\end{tabular}
\caption{NUHM2 input parameters and masses in GeV units for the two {\it
radiatively-driven natural SUSY} benchmark points introduced in the text.
We take $m_t=173.2$~GeV}
\label{tab:bm}
\end{table}
%

\section{Sparticle production and decay at a higgsino factory}
\label{sec:prod}

\subsection{Sparticle production}

In Fig.~\ref{fig:xsec}, we show sparticle and Higgs boson production rates for
unpolarized beams at the ILC versus $\sqrt{s}$ for the ILC1 benchmark
point.  Rates were computed at leading order, with leading order
spectra, using formulae from Ref.~\cite{bbkmt}. Also shown for
comparison is the rate for muon pair production.  We see that around
$\sqrt{s}\sim 220-230$~GeV, the threshold for production of $Zh$,
$\tw_1^\pm\tw_1^\mp$ and $\tz_1\tz_2$ is crossed so that production
rates rise rapidly.  Whereas one might expect ILC at these energies to
be a Higgs boson factory, we see that  ILC would also be a higgsino factory,
where the higgsino pair production rates exceed $Zh$ production by
factors of 5-10!  While the higgsino decay debris may be  too soft to be
picked out above SM backgrounds at LHC, the clean environment of a linear
$e^+e^-$ collider allows for straightforward discovery, as
discussed in Sec.~\ref{sec:det}. Thus, while radiatively-driven
natural SUSY might well elude LHC searches, it cannot elude searches at
ILC provided that $\sqrt{s}> 2m(higgsino)$.
\FIGURE[tbh]{
\includegraphics[width=15cm,clip]{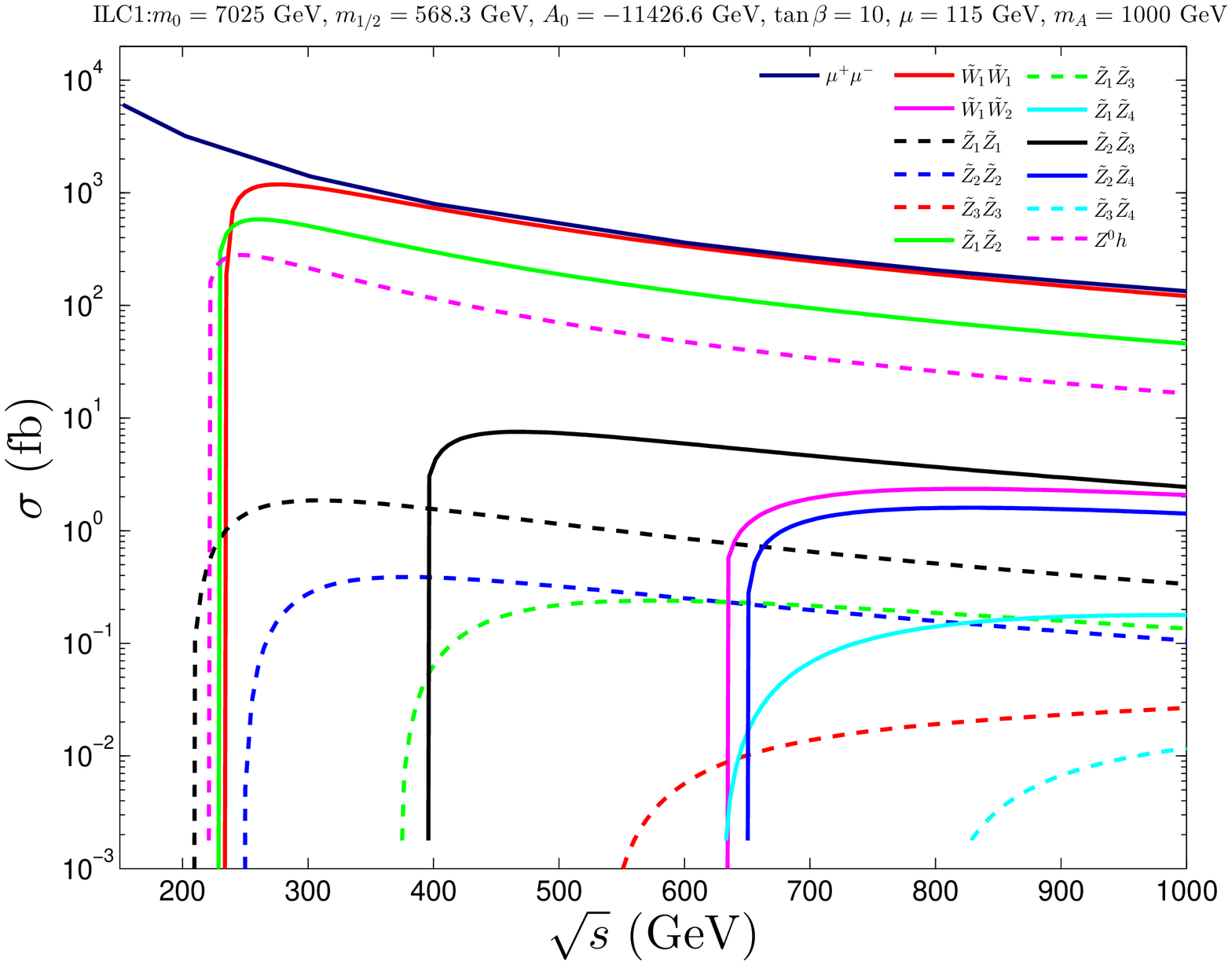}
\caption{Sparticle production cross sections vs. $\sqrt{s}$ for
  unpolarized beams at an $e^+e^-$ collider for the ILC1 benchmark point
  listed in Table \ref{tab:bm}.}
\label{fig:xsec}}

While the reactions $e^+e^- \to \tw_1^+\tw_1^-$ and $e^+e^- \to \tz_1\tz_2$ will be the
first sparticle production processes accessed at ILC250, 
the discovery prospects do not end there.
As $\sqrt{s}$ is increased beyond $2m(higgsino)$, further thresholds will be
passed, including those for $\tz_2\tz_3$, $\tw_1\tw_2$ and $\tz_2\tz_4$
production. 
These occur at somewhat lower but still measureable rates. 
Even reactions with much lower production rates -- such as
$\tz_2\tz_2$, $\tz_1\tz_3$, $\tz_3\tz_3$ and $\tz_3\tz_4$ -- might
ultimately be detectable, depending on the machine energy and integrated
luminosity that is ultimately attained.

In Fig.~\ref{fig:pol}, we show the $\tw_1^+\tw_1^-$ and  $\tz_1\tz_2$
production rates for the ILC1 benchmark case at $\sqrt{s}=250$~GeV,
but as a function of the electron beam polarization $P_L(e^-)$.  Here
and in the rest of the paper, we assume that the positron beam is
unpolarized, {\it i.e.} $P_L(e^+)=0$.  Whereas $\tw_1^+\tw_1^-$
production has the largest rate for unpolarized beams ($P_L(e^-)=0$),
for the case of right polarized electron beam, $\sigma (\tw_1^+\tw_1^-)$
diminishes by a factor of about 4 and instead $\sigma (\tz_1\tz_2)$,
which is much less sensitive to beam polarization, is dominant.  The
comparable rates (within an order of magnitude) for both both chargino
and neutralino pair production (solid curves), together with the
relatively mild polarization is characteristic of the production of
higgsino-like charginos and neutralinos. For wino-like gauginos in the
kinematically accessible range, chargino production would occur at a
high rate, but neutralino pair production would be strongly suppressed
because $SU(2)_L\times U(1)_Y$ gauge symmetry forbids couplings of the
$Z$ and $\gamma$ to both binos and (neutral) winos.\footnote{This
assumes that the selectron is heavy so that neutralino production via
$t$-channel selectron production is negligible. Neutralino production
via $t$-channel selectron exchange also  yields a large rate for
$\tz_2\tz_2$ production, so should be readily distinguishable since
there would be events also in the $4\ell$, $2\ell 2j$ and $4j +\eslt$
channels. The angular distributions of the neutralinos will
also be different if $t$-channel exchange contributions are significant.}
This can be seen in the dashed curve in Fig.~\ref{fig:pol} which shows
the cross section for $\tw_1\tw_1$ production for the ILC1 NUHM2 model
point except that $m_{1/2}$ and $\mu$ are now chosen so that the weak
scale values of $M_2$ and $\mu$ are essentially exchanged. In this case,
the masses of the wino-like $\tw_1$ and $\tz_2$ is about the same as for
the higgsinos of the ILC1 point.  The neutralino-pair production cross
sections for this wino-like case are below 0.1~fb and do not show up in
this frame.  This observation will be important in Sec.~\ref{sec:det} where we
describe our analysis.  The polarization dependence of the chargino pair
production cross section provides an independent handle that may enable
us to argue the higgsino-like nature of the charginos of the ILC1
point. For a right-handed electron beam the amplitude for charged wino
pair production is suppressed by a factor of $M_W^2/s$ relative to that
for charged higgsino pair production, accounting for the strong drop of
the dashed curve at $P_L(e^-) = -1$.

%
\FIGURE[tbh]{
\includegraphics[width=13cm,clip]{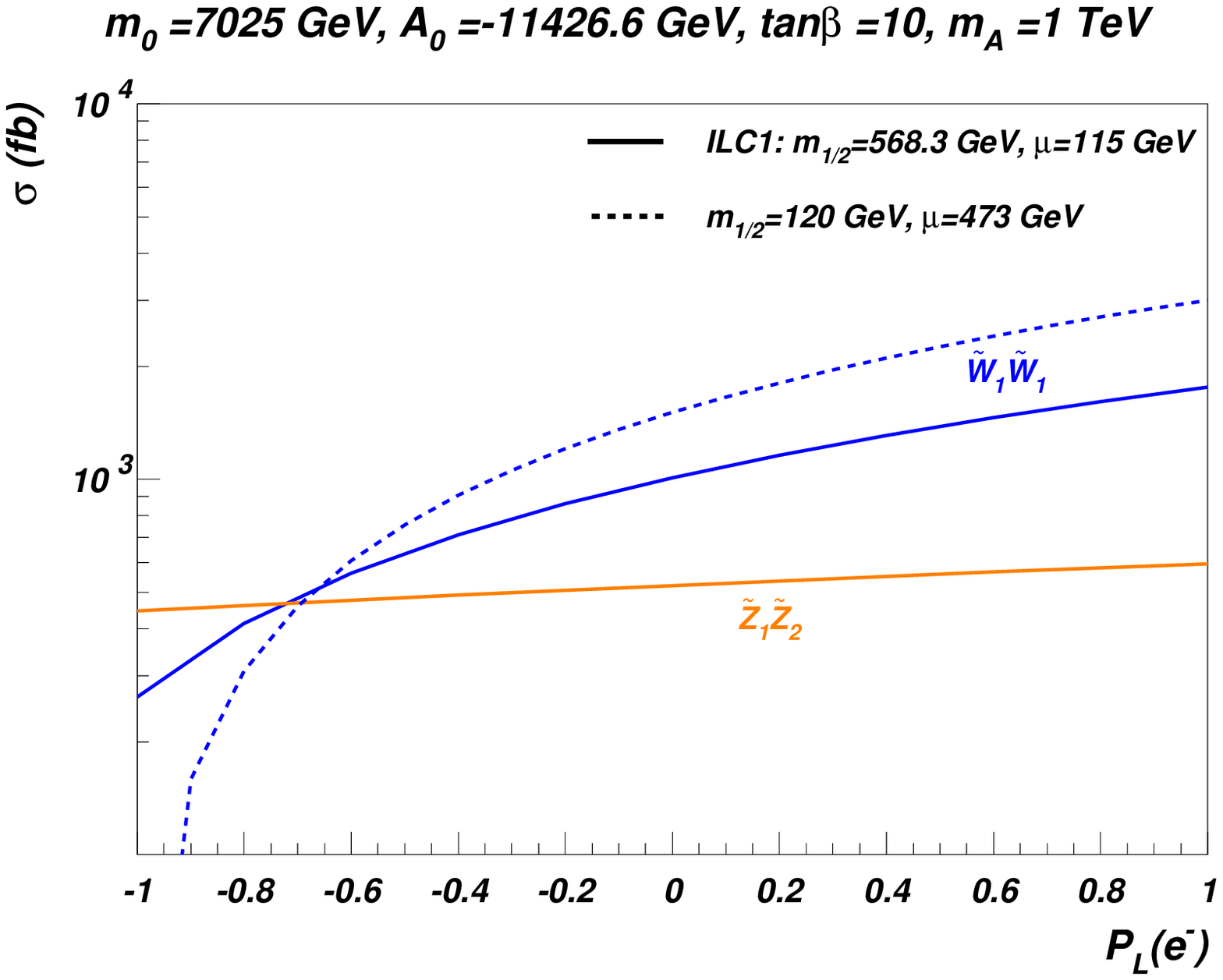}
\caption{Sparticle production cross sections vs. $P_L(e^-)$ at an
$e^+e^-$ collider for the ILC1 benchmark point with
$\sqrt{s}=250$~GeV. The positrons are taken to be unpolarized. For
comparison, we show a point with a wino-like chargino of similar
mass. For the wino-like case with $m_{1/2}=120$~GeV, then the $\sigma
(e^+e^-\to \tz_1\tz_2)\sim 0.1$~fb, while $\sigma(\tz_2\tz_2)$ is even
smaller, and so is far below the cross section values shown.  }
\label{fig:pol}}

\subsection{Higgsino decays}

Since the inter-higgsino mass gaps are so small, for the case of RNS one
expects the following three-body decays to be dominant: 
\bea
\tw_1^-&\to & \tz_1 f\bar{f}'\;,\\ \tz_2& \to & \tz_1 f\bar{f}\;, \eea 
where the $f$ stand for SM fermions.  Despite the larger phase space
suppression for the three body decays,
the branching fraction for the loop decay $\tz_2\to \tz_1\gamma$ is
still small because of the large $Z\tz_1\tz_2$ coupling \cite{loop}.
Moreover, squark and slepton masses are expected very large within the
RNS framework, and  the $\tw_1$ and $\tz_2$ three-body
decay amplitudes are dominated by $W^*$ and $Z^*$ exchange,
respectively. The branching fractions into specific modes will thus
closely follow the corresponding $W$ and $Z$ decay branching fractions,
{\it i.e.}  we obtain $B(\tw_1^-\to \tz_1 e^-\bar{\nu}_e) \simeq 11\%$,
$B(\tz_2 \to e^+e^-\tz_1) \simeq 3\%$, {\it etc.}.

\section{SUSY event generation} 
\label{sec:evgen}
%

Within the RNS framework, higgsino pair production at the ILC will be
signalled by events with {\em low visible energy} from the relatively
soft daughter leptons and jets from $\tw_1$ and $\tz_2$ decays, and
modest $\eslt$. We need, therefore, to pay particular attention to
SM sources of low visible energy events. Since the bulk of $2\to 2$
events lead to large visible energy, the most important backgrounds
come from two photon processes, $e^+e^- \to e^+e^- f\bar{f}$, where the
energetic final state electron and positron are lost down the
beam-pipe, and the visible energy in the detector arises only from 
$f$ and $\bar{f}$. 

We use ISAJET v7.84 for our SUSY event simulation as well as simulation
of $2 \to 2$ and $\gamma\gamma$-induced SM backgrounds.
The  $2\to 2$ SM background sources include
\be
e^+e^-\to f\bar{f},\ W^+W^-\ \ {\rm and}\ \ Z^0Z^0\;,
\label{eq:2to2}
\ee
while $\gamma\gamma$ processes include, 
\be
\gamma\gamma\to \tau^+\tau^-,\ \ c\bar{c}\ \ {\rm and}\ \ b\bar{b}.
\label{eq:2to4}
\ee
The reaction $e^+e^-\to Zh$ is included in the signal sample, but plays
no role here.

We use the Isajet toy calorimeter covering
$-4<\eta <4$ with cell size $\Delta\eta\times\Delta\phi =0.05\times 0.05$.
Energy resolution for electromagnetic and hadronic depositions is taken to be
$\Delta E_{em}/E_{em} =0.15/\sqrt{E_{em}}\oplus 1\%$ and 
$\Delta E_h/E_h=0.5/\sqrt{E_h}/E_h\oplus 2\%$, respectively (where
$\oplus$ denotes addition in quadrature).
Jets are found using fixed cones of size
$R=\sqrt{\Delta\eta^2+\Delta\phi^2} =0.6$ using the ISAJET routine
GETJET (modified for clustering on energy rather than transverse energy).
Clusters with $E>5$~GeV and $|\eta ({\rm jet})|<2.5$ are labeled as jets.
Muons and electrons are classified as isolated if they have $E>5$~GeV,
$|\eta (\ell )|<2.5$, and the visible activity within a cone of $R=0.5$
about the lepton direction is less than 
$min({E_{\ell}\over 10},\ 1\ {\rm GeV})$. 

Our production reactions are run using electron PDFs which include a
convolution of bremsstrahlung~\cite{brem} and beamstrahlung~\cite{beam}
contributions as described in Ref.~\cite{tadas2}.  For $\sqrt{s}=250$~GeV, 
we use beamstrahlung parameter $\Upsilon =0.02$ and for
$\sqrt{s}=340$~GeV, we use $\Upsilon =0.03$.  For both cases, we use
the beam bunch length $\sigma_z =0.3$~mm~\cite{tdr3}.

For processes with low visible energy, the two-photon processes 
$\gamma\gamma\to f\bar{f}$ can be the most relevant.
These processes also give rise to  $\eslt$ when the decay products of $f$
include neutrinos. For our analysis we therefore include only 
\bi
\item $\gamma\gamma\to\tau^+\tau^-,\ c\bar{c}$ and $b\bar{b}$
\ei
contributions from Isajet using a photon PDF which again
includes a beam/bremsstrahlung convolution\cite{drees,tadas2}.

\section{Detection of higgsinos at a higgsino factory}
\label{sec:det}
%

\subsection{Benchmark ILC1 at $\sqrt{s}=250$~GeV}

We begin by first discussing higgsino pair production for the ILC1
benchmark point with $\sqrt{s}=250$~GeV, the nominal turn-on energy of
the ILC.  Once
threshold for pair production is passed, then the two higgsino pair
production reactions occurring at the highest rates are 
\bi
\item $e^+e^-\to \tw_1^+\tw_1^-\to (f\bar{f}'\tz_1)+(F\bar{F}'\tz_1$)\ \  and 
\item $e^+e^-\to \tz_1\tz_2\to \tz_1+(f\bar{f}\tz_1)$ 
\ei 
where $f$ and $F$ are SM fermions. For models where $|\mu|\ll M_{1,2}$, the two
lightest neutralinos are well approximated by $(\th_u\pm \th_d)\over
\sqrt{2}$, and the coupling of $Z$ to $\tz_1\tz_1$ and $\tz_2\tz_2$
pairs is dynamically suppressed~\cite{wss}. Thus, though the phase space
for $\tz_2\tz_2$ production is qualitatively similar to that for
$\tz_1\tz_2$ production, $\sigma(\tz_2\tz_2)$ is much smaller in the RNS
framework: see Fig.~\ref{fig:xsec}.

Since $m_{\tz_1}$ is only slightly smaller than $m_{\tw_1,\tz_2}$, most
of the collision energy ends up in the rest mass $2m_{\tz_1}$ of the
LSPs, and the visible final state fermions are relatively soft.  To
illustrate this, we show in Fig.~\ref{fig:Evis} the visible
energy distribution expected at ILC250 for benchmark ILC1.  From the
figure, we see the bulk of SM background from $e^+e^-$ annihilation
 processes (green curve) in (\ref{eq:2to2}) peaks around $E_{vis}\sim
250$~GeV, with some spillover to higher values due to detector energy
mis-measurement. A continuous $E_{vis}$ tail occurs at lower values due
to production of $WW$, $ZZ$, $b\bar{b}$ {\it etc.} where substantial
energy is lost due to decays to neutrinos. There are also two small
bumps at $E_{vis}\sim 140$~GeV and $250$~GeV arising from $Zh$ production
(blue curves). The 140~GeV bump occurs due to $e^+e^-\to Zh\to
(\nu\bar{\nu})+(b\bar{b})$.  The SUSY signal distribution is depicted by
the bounded (red) histogram with $E_{vis}\sim 0-40$~GeV is already
well-separated from the $2\to 2$ SM backgrounds.  However, as
anticipated, backgrounds from the $2\to 4$ processes, $\gamma\gamma\to
c\bar{c},\ b\bar{b}$ and $\tau\bar{\tau}$, shown as the black histogram
overwhelm the signal by a factor of $\sim 100-1000$.  
\FIGURE[tbh]{
\includegraphics[width=12cm,clip]{etot_250.eps}
\caption{Distribution in visible energy measured in $e^+e^-$ events at
$\sqrt{s}=250$~GeV for ILC1 signal and SM backgrounds from
$e^+e^-$ and $\gamma\gamma$ collisions.
We take beamstrahlung parameters $\Upsilon =0.02$ and $\sigma_z=0.3$~mm.
}
\label{fig:Evis}}

To select out signal events, we first require:
\bi
\item $20$~GeV $<E_{vis}<50$~GeV.  
\ei 
The $\gamma\gamma$ background yields mainly soft visible energy events
with a tail extending to higher values. To differentiate signal from
this background, we plot in Fig.~\ref{fig:ETM} the missing transverse
energy distribution $d\sigma /d\eslt$ after the visible energy cut. We
see that the $\gamma\gamma$ background falls very rapidly since $\eslt$
occurs mainly due to neutrinos from the decays of the relatively light
$c,b$ and $\tau$, and the signal emerges from $\gamma\gamma$ background
if we require $\eslt > 10$~GeV.  The bulge of events with low $E_{vis}$
but modest $\eslt$ would herald the discovery of new physics.  
This also explains why we did  not
include $\gamma\gamma \to f\bar{f}$ processes with $f=e,\mu,u,d,s$ in 
our analysis. These
yield  back-to-back events in the transverse plane, with
essentially no $\eslt$, and are efficiently eliminated by a $\eslt$ cut.
Thus, for
our new physics event sample, we will also require 
\bi
\item $\eslt >10$~GeV.
\ei
\FIGURE[tbh]{
\includegraphics[width=12cm,clip]{ptm_250.eps}
\caption{Distribution of missing transverse energy from $e^+e^-$ collisions at
$\sqrt{s}=250$~GeV for ILC1 signal along with SM background
from $e^+e^-$ and $\gamma\gamma$ collisions.
We require $20$~GeV$<E_{vis}<50$~GeV.
We take beamstrahlung parameters $\Upsilon =0.02$ and $\sigma_z=0.3$~mm.
}
\label{fig:ETM}}

To understand the expected event topologies, we examine the multiplicity
of isolated leptons and identied jets.  These distributions are shown in
Fig.~\ref{fig:nlepjet}.  We see that the most lucrative signal channels
from the perspective of the signal to background ratio appears to
be the $n(\ell)=0$ and $n(jet)=1-3$ bins to which neutralino and chargino 
production can contribute. To cleanly separate chargino and neutralino
contributions so that each particle can be studied in detail, it is
also useful to examine other channels. 

Before turning to this, we note that the observation of an excess above
SM in the multi-jet plus multi-lepton channels, if interpreted in terms
of weak scale SUSY, would suggest the production of charginos and
neutralinos. The small energy release in these events would point to a
small mass gap between the parent particles and the undetected LSP. In
the simplest models with gaugino mass unification, this would indicate
the production of higgsino-like states, with $|\mu| \ll m_{1/2}$ where
$\tw_1$, $\tz_2$ and $\tz_1$ are roughly degenerate, and the bino and
winos are substantially heavier. However, it is also possible that such
events may arise from wino pair production in models with heavy
higgsinos, and a bino only slightly lighter than the wino-states. It
should, however, be possible to distinguish between these possibilities
since, as we mentioned in our discussion of Fig.~\ref{fig:pol}, with
unpolarized beams neutralino production is smaller than 0.1~fb ({\it
i.e.} three orders of magnitide below expectations for higgsino-like
neutralinos) in the latter case. Moreover, the chargino signal from the
production of wino-like charginos will reduce much more sharply for
right-handed electron beams than for higgsino-like charginos. It should
thus be possible to unambiguously conclude that the signal is from
higgsino-like, and not gaugino-like super-partners. 

With this in mind, we
turn to strategies that will help us to obtain essentially pure samples
of chargino and of neutralino events for the ILC1 point under
examination.
\FIGURE[tbh]{
\includegraphics[width=10cm,clip]{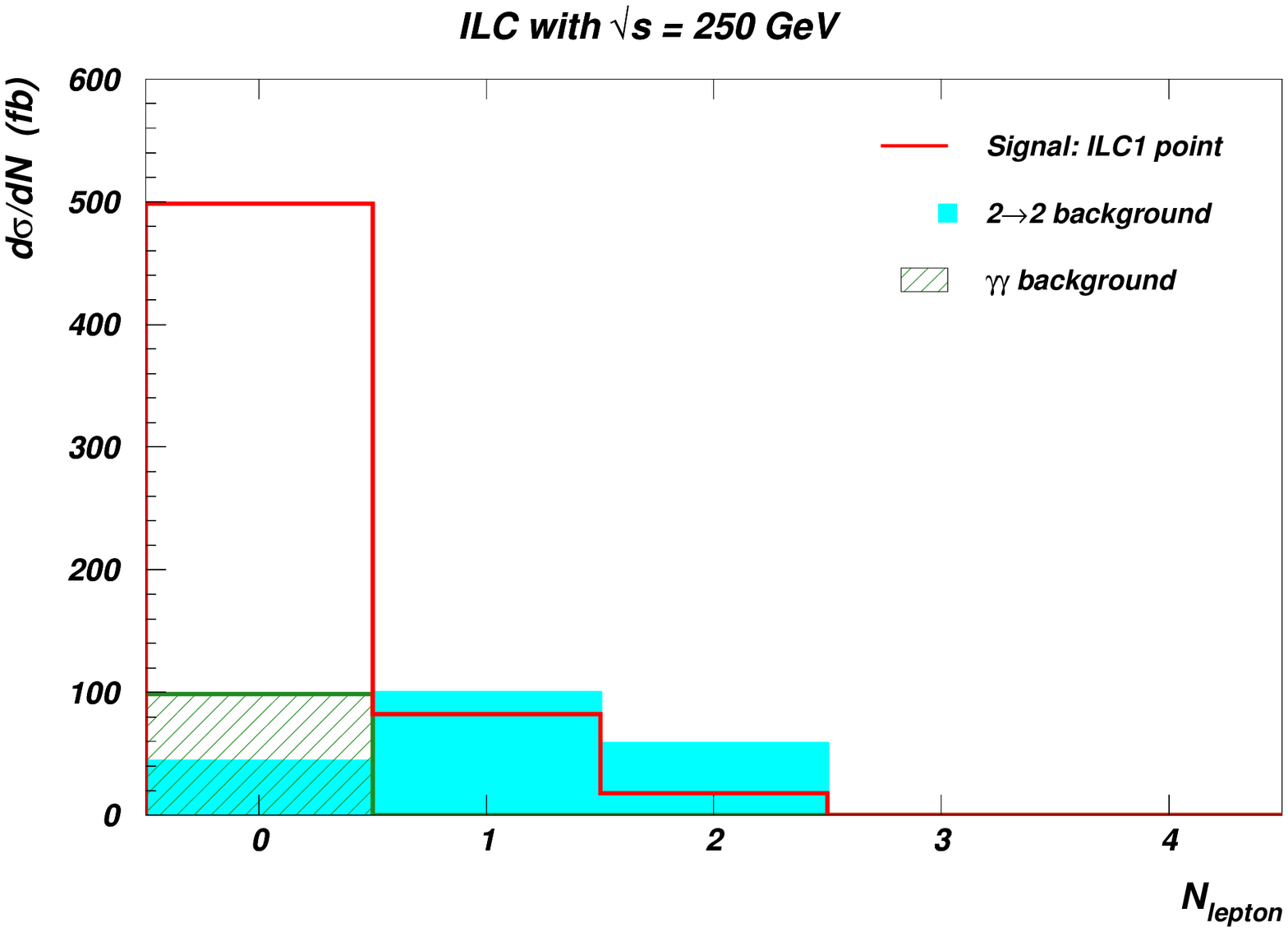}
\includegraphics[width=10cm,clip]{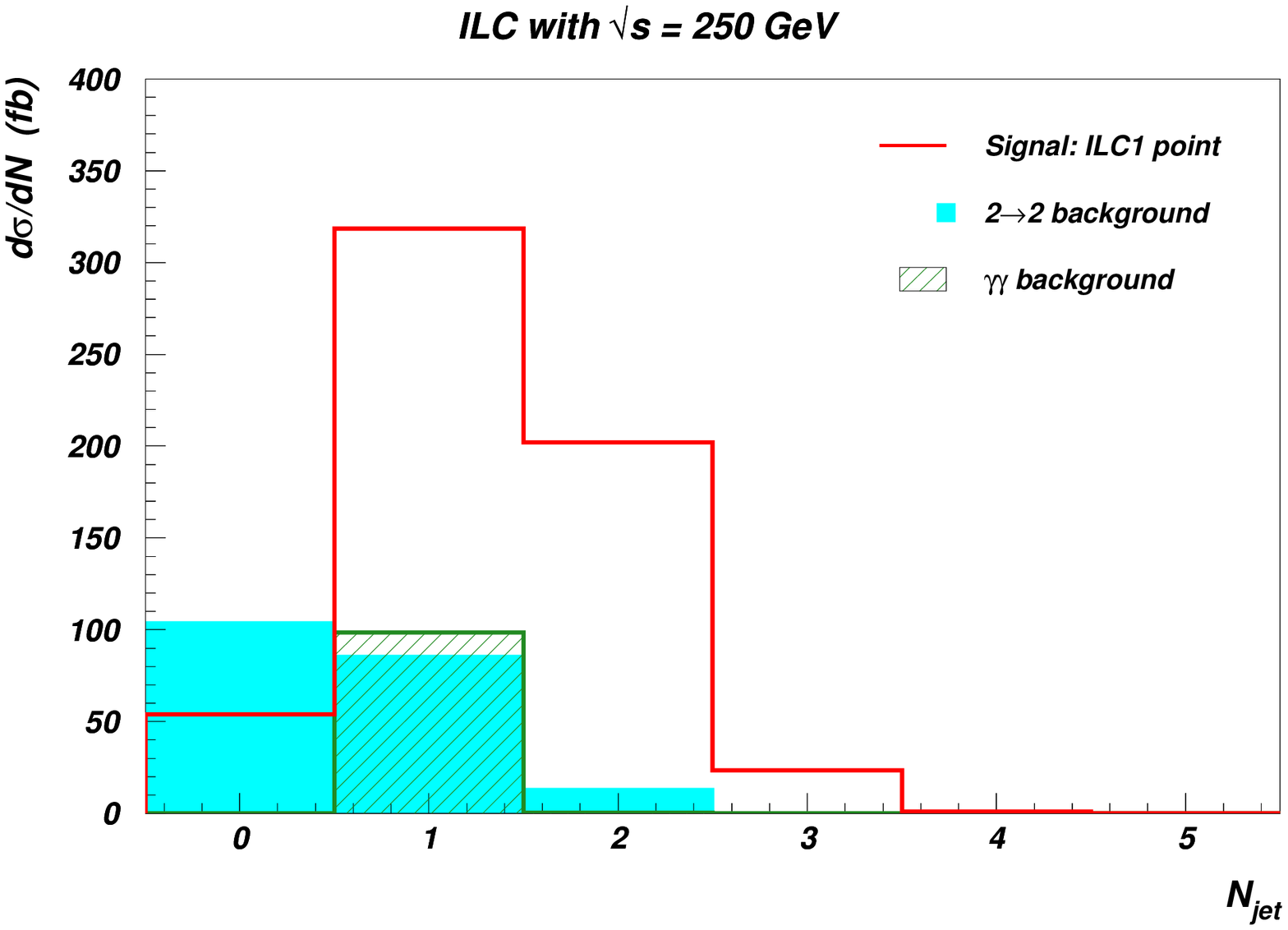}
\caption{Distribution of {\it a}) isolated lepton multiplicity and {\it
b}) jet multiplicity from $e^+e^-$ collisions at $\sqrt{s}=250$~GeV for
higgsino signals from the ILC1 case study
along with corresponding SM backgrounds from $e^+e^-$ and
$\gamma\gamma$ collisions.  We require $20$~GeV$<E_{vis}<50$~GeV and
$\eslt >10$~GeV.  We take beamstrahlung parameters $\Upsilon =0.02$ and
$\sigma_z=0.3$~mm.  }
\label{fig:nlepjet}}

\subsubsection{Chargino pair production}
\label{sssec:w1w1}

To select out a nearly pure sample of chargino pair events where the jets all
arise from the same chargino, we will first select events with the
$E_{vis}$ and $\eslt$ cuts introduced above, but also require 
\bi
\item $n(\ell)=1$
\ei
 and
\bi 
\item $n(jet)=2$.
\ei
After these requirements, we are left with a signal cross section of 6.43~fb. 
Just one background event passes cuts, leading to $\sigma_{BG}\sim 0.018$~fb, 
{\it i.e.} we have a nearly pure sample of chargino pair events, 
where one chargino decays leptonically and the other decays hadronically.

We show a scatter plot of these selected events in the $E(jj)$
vs. $m(jj)$ plane in Fig.~\ref{fig:ejjmjj}.
\FIGURE[tbh]{
\includegraphics[width=12cm,clip]{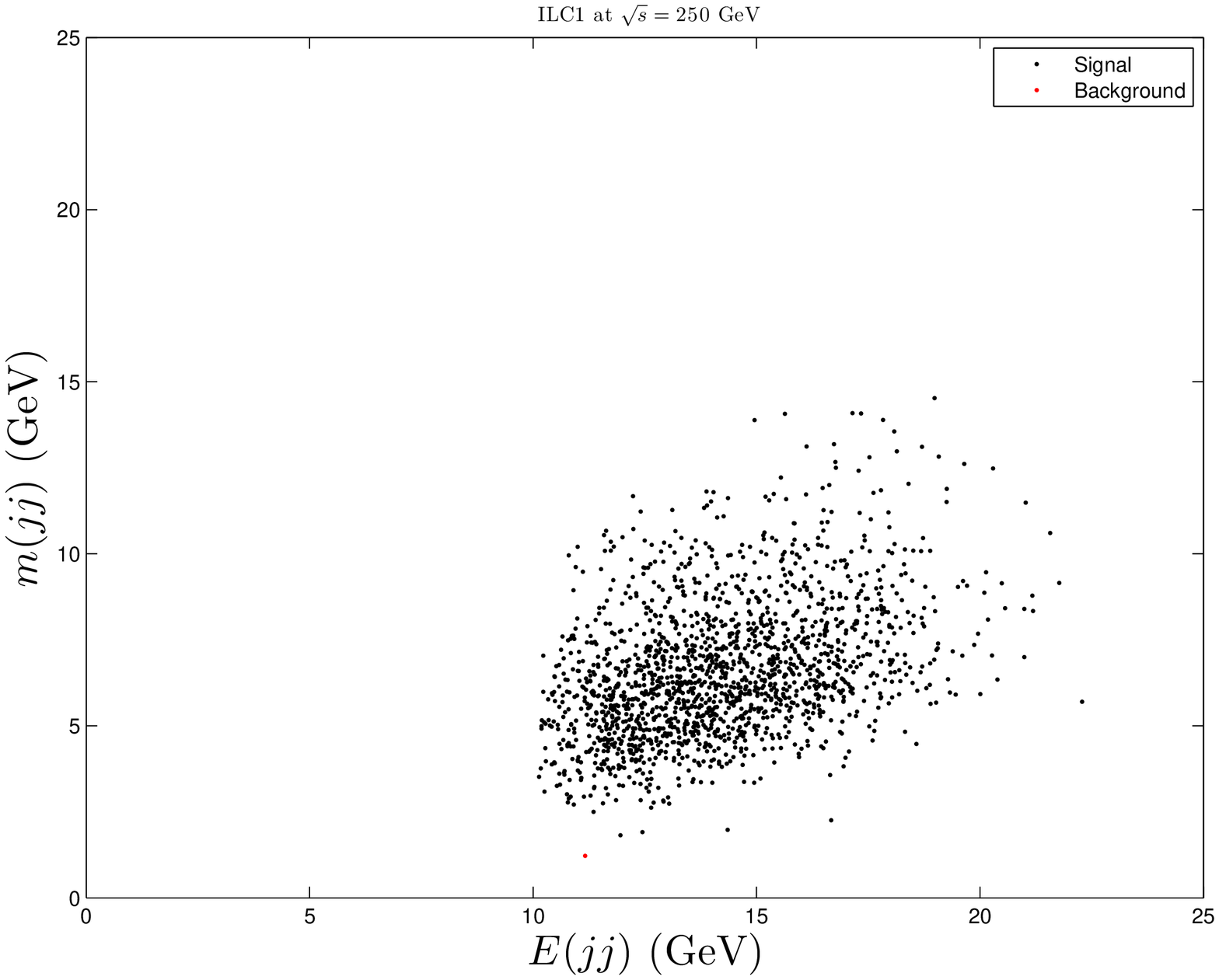}
\caption{Scatter plot in the $E_{jj}$ vs. $m(jj)$ plane for $1\ell
+2-jets$ events from the ILC1 point in $e^+e^-$ collisions at
$\sqrt{s}=250$~GeV. We require $20$~GeV$<E_{vis}<50$~GeV and $MET>10$
GeV.  We take beamstrahlung parameters $\Upsilon =0.02$ and
$\sigma_z=0.3$~mm.  }
\label{fig:ejjmjj}}
The $m(jj)$ distribution is expected to be bounded from above by
$m_{\tw_1}-m_{\tz_1}=14.6$~GeV up to energy mis-measurement corrections;
this cut-off is seen in Fig.~\ref{fig:ejjmjj}, from which it is apparent the
$m_{\tw_1}-m_{\tz_1}$ mass gap is $\sim 15$~GeV.

The sparticle masses $m_{\tw_1}$ and $m_{\tz_1}$  can be obtained 
from fits of the $E(jj)$ data distribution\cite{jlc,bmt,tadas1} 
to various expected theory distributions which vary depending on
$m_{\tw_1}$ and $m_{\tz_1}$. The lower endpoint of $E(jj)$ is determined
largely by our $E(j)>5$~GeV jet requirement but the upper endpoint is
quite sensitive to $m_{\tw_1}$ and $m_{\tz_1}$ values.

To assess the precision which can be attained, we generate a synthetic
``data'' set assuming 100~fb$^{-1}$ of integrated luminosity, along with
expected statistical error bars.  We also generate theory sample of
distributions run over a large grid of $\mu$ and $m_{1/2}$ points (which
yields a corresponding grid of $m_{\tw_1}$ and $m_{\tz_1}$ points) where
each theory sample is run with 10 times the statistics of data. Our
analysis ignores any sensitivity to other parameters, and implicitly
assumes that we can distinguish between higgsino- and wino-like chargino
events (which should be possible as noted just before the start of
Sec.~\ref{sssec:w1w1}).  We then compare the $E(jj)$ ``data'' distribution
 (with 1~GeV bins) to these theory templates and obtain the values of
$\chi^2$ between the ``data'' and the theory.  We fix the
normalization of theory curves to match ``data'' so that we are fitting
only the shape of the distribution. To obtain the $\chi^2$, we add the 
appropriately weighted statistical errors for the theory and data sets
in quadrature.

This procedure enables us to obtain a grid of $\Delta\chi^2
=\chi^2-\chi^2_{min}$ values in the $m_{\tw_1}-m_{\tz_1}$ plane.  The
reader should keep in mind that our theory calculation  is also subject
to statistical fluctuations that will be reflected in the distribution 
of $\Delta\chi^2$ values. To enable the reader to personally assess the
reliability of the computation, we show in Fig.~\ref{fig:chi2}{\it a} these
$\Delta\chi^2$ values binned by $\Delta\chi^2 < 2.3$ ($1\sigma$ CL),
$\Delta\chi^2 < 4.6$ (90\% CL) and $\Delta\chi^2 > 4.6$. We also show
the corresponding $1\sigma$ and 90\% CL error ellipses that we obtain as 
conservative fits to the $\Delta\chi^2$ data.
\FIGURE[tbh]{
\includegraphics[width=12cm,clip]{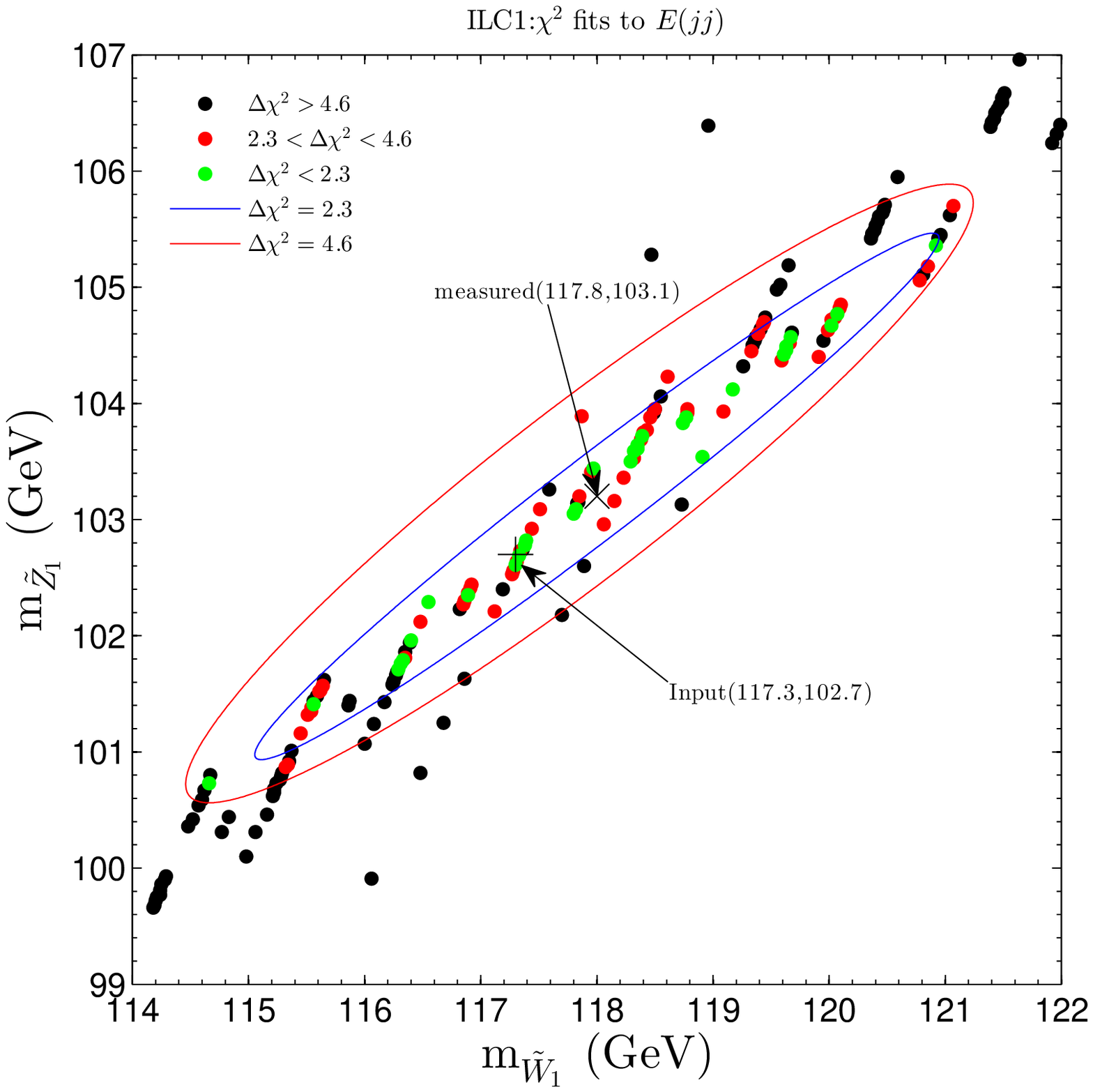}
\includegraphics[width=10cm,clip]{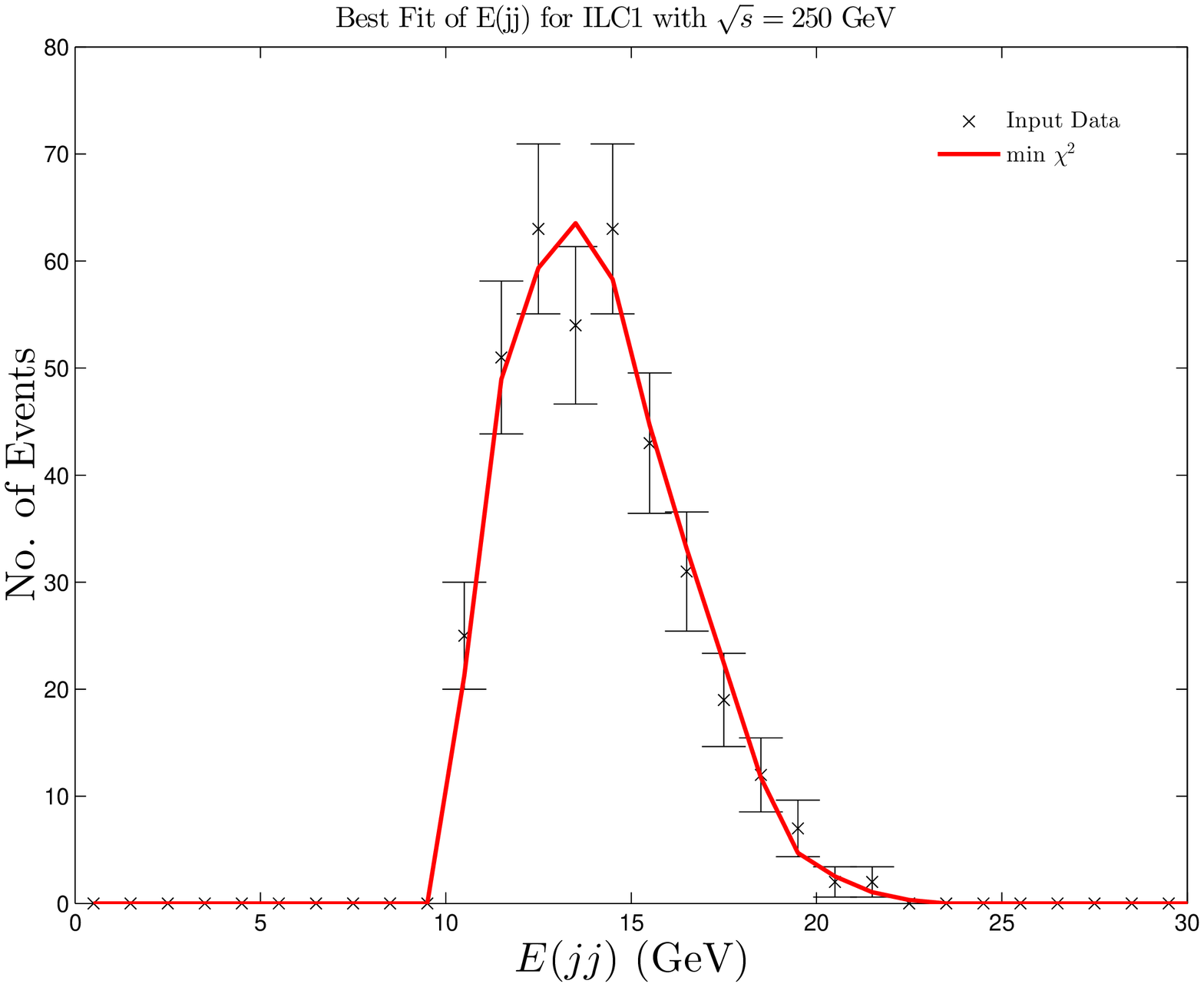}
\caption{In {\it a}), we show values of $\Delta\chi^2$ found from
matching 100~fb$^{-1}$ of ILC1 ``data'' to various ``theory''
distributions generated from a scan over $\mu\ vs.\ m_{1/2}$ space. Each
``theory'' point is run with ten times the events contained in the
``data'' distribution.  We also show fitted error ellipses corresponding
to $1\sigma$ and 90\% CL measurements.  In
{\it b}), we show the distribution in $E(jj)$ from 100~fb$^{-1}$ of
``data'' along with best fit distribution.  }
\label{fig:chi2}}
>From these error ellipses, we find  that the 2-3\% mass  measurements 
\bi
\item $m_{\tw_1}=117.8\pm 2.8$~GeV ($1\sigma$),
\ei
 and also,
\bi
\item $m_{\tz_1}=103.1\pm 2.2$~GeV ($1\sigma$), \ei 
should be possible
for the ILC1 point.  The  synthetic ``data'',
together with statistical error bars corresponding to an integrated
luminosity of 100~fb$^{-1}$, are
shown in Fig.~\ref{fig:chi2}{\it b}) along with the best fit
distribution shown as the solid curve.

Here, we
remark that if instead the time is taken to perform various total cross
section measurements around the higgsino pair threshold -- which will
require a much higher integrated luminosity investment at several
$\sqrt{s}$ values \cite{threshold} -- even better precision on the
masses may be expected.

\subsubsection{Neutralino pair production}
\label{sssec:z1z2}

For the case of $\tz_1\tz_2$ pair production, we examined events where
$\tz_2\to q\bar{q}\tz_1$ that yield an  $n(\ell)=0,\ n(j)=2$ sample as well as
events where $\tz_2 \to \ell^+\ell^-\tz_1$, for which $n(\ell)=2$ and $n(j)=0$. 
While one might expect the dijet sample to yield more events due to
the large $\tz_2\to\tz_1 q\bar{q}$ branching fraction, in fact we find
after cuts that the $\ell^+\ell^-$ sample is larger.  This is because
frequently the two possible quark jets merge to yield only a single
resolvable jet given our jet finding algorithm, or else one of the
possible jets becomes too soft or too forward to be identified.

For the opposite-sign/same flavor (OS/SF) dilepton signal that we focus
on, we use a polarized electron beam with $P_L(e^-)=-0.9$ since this 
helps to reduce potential backgrounds from $WW$ production, and also limits 
contamination from chargino production to around 10\%. 
We then require: 
\bi
\item exactly 2 OS/SF isolated leptons with no jets, 
\item $E_{vis}<35$~GeV,
\item transverse plane angle between the two leptons
$\Delta\phi(\ell^+\ell^-)< \pi /2$.  
\ei 
After these cuts, the
$\gamma\gamma$ background is eliminated but some SM background -- mainly
$WW$ production -- remains.  
The $E(\ell^+\ell^- )$ distribution after these cuts is shown in
Fig.~\ref{fig:Ell}.  This leaves a OS/SF dilepton signal of 19.55~fb
while SM background is $0.44$~fb. 
We have checked that the signal has a negligible contribution from
chargino production.
\FIGURE[tbh]{
\includegraphics[width=12cm,clip]{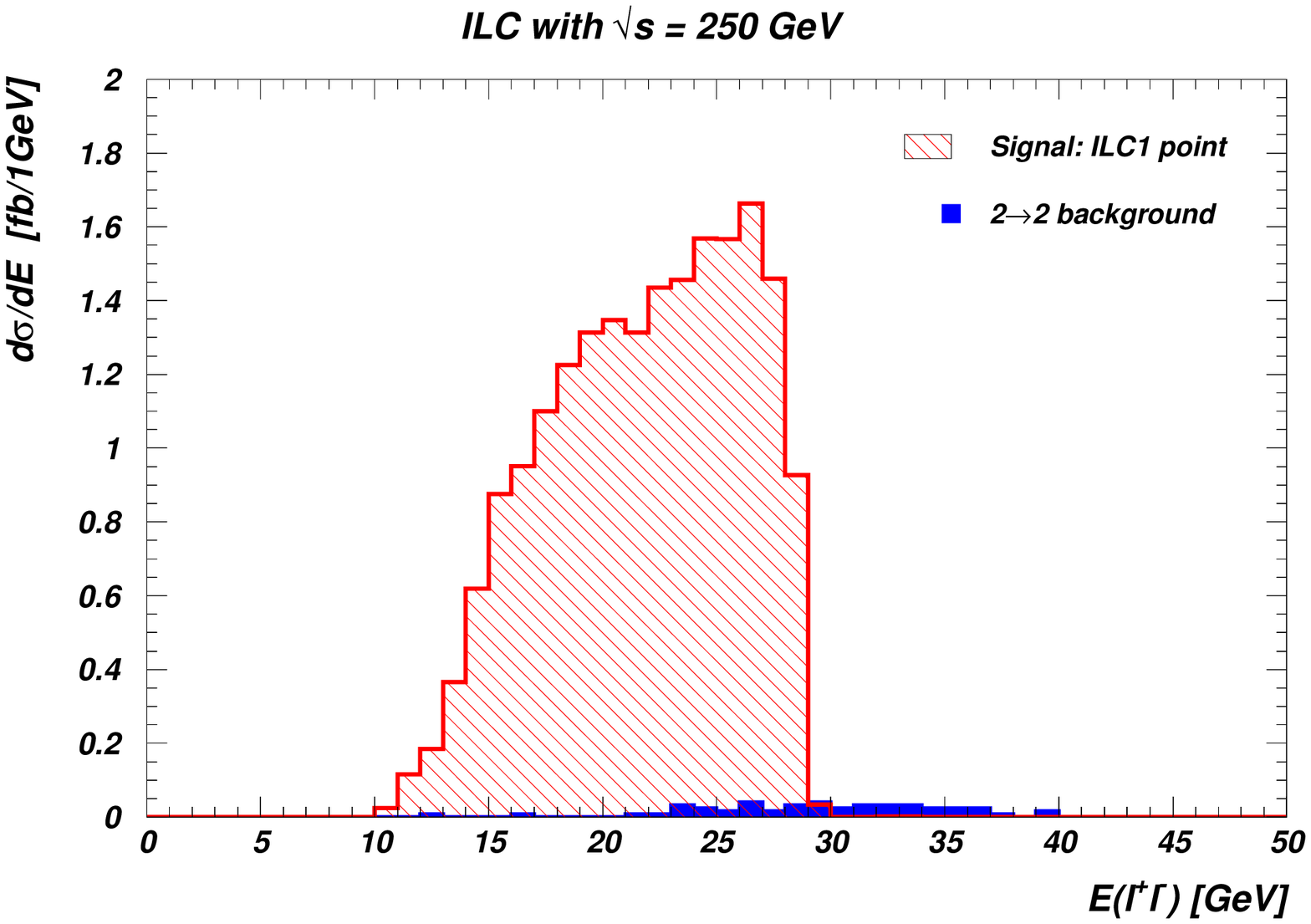}
\caption{Distribution in $E(\ell^+\ell^- )$ from 100~fb$^{-1}$ of
``data'' of OS/SF dilepton events with $P_L(e^-)=-0.9$ from the signal
for the ILC1 case, and from SM background (which only comes from $2\to 2$
processes), after the $E_{vis}$ and $\Delta\phi(\ell\ell)$ cuts
discussed in the text.}
\label{fig:Ell}}

Armed with the clean sample of OS/SF dilepton signal events from
essentially $\tz_1\tz_2$ production, we next examine the $m(\ell^+\ell^-
)$ distribution for the ILC1 case.  We expect that this distribution is
kinematically bounded from above by the $m_{\tz_2}-m_{\tz_1}$ mass
difference and relatively insensitive to the absolute masses of the
particles.  We use the theory templates generated with 10 times the
statistics, as described in Sec.~\ref{sssec:w1w1} to obtain a map of
$\chi^2$ vs. $m_{\tz_2}-m_{\tz_1}$, shown by the jagged (black) line in
Fig.~\ref{fig:chi2_gap}{\it a}. As before, we fit to the shape, allowing
the normalization to float. While statistical fluctuations do
contribute to the jaggedness, we have checked that the points with the
largest $\chi^2$ values come from theory templates where the mass scale
of the neutralinos is very different. Also shown in the figure
is a parabolic fit to the values of $\chi^2$. We 
see that the mass gap should be
measureable at the percent level: 
\bi
\item $m_{\tz_2}-m_{\tz_1} =21.0\pm 0.2$~GeV  ($1\sigma$). 
\ei
The best fit line and the dilepton mass ``data'' used to obtain the fit
 are shown in the lower frame in the figure. We see that the shape of
 this mass distribution is indicative of an opposite sign of the $\tz_1$
 and $\tz_2$ mass eigenvalues~\cite{kadala}, completely compatible with
 expectation~\cite{kn} from the decay of a higgsino-like $\tz_2$ to a
 higgsino-like $\tz_1$.
\FIGURE[tbh]{
\includegraphics[width=11cm,clip]{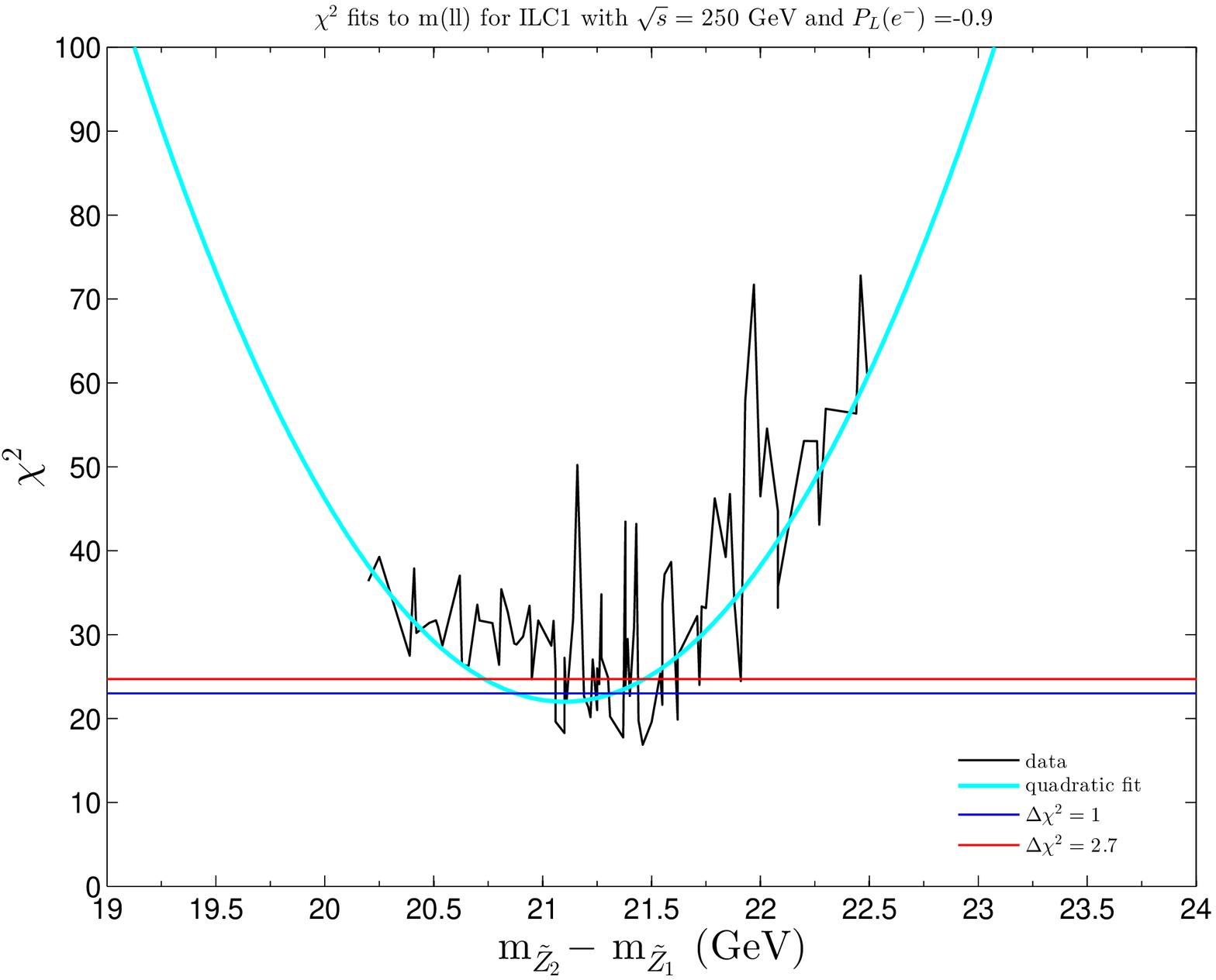}
\includegraphics[width=11cm,clip]{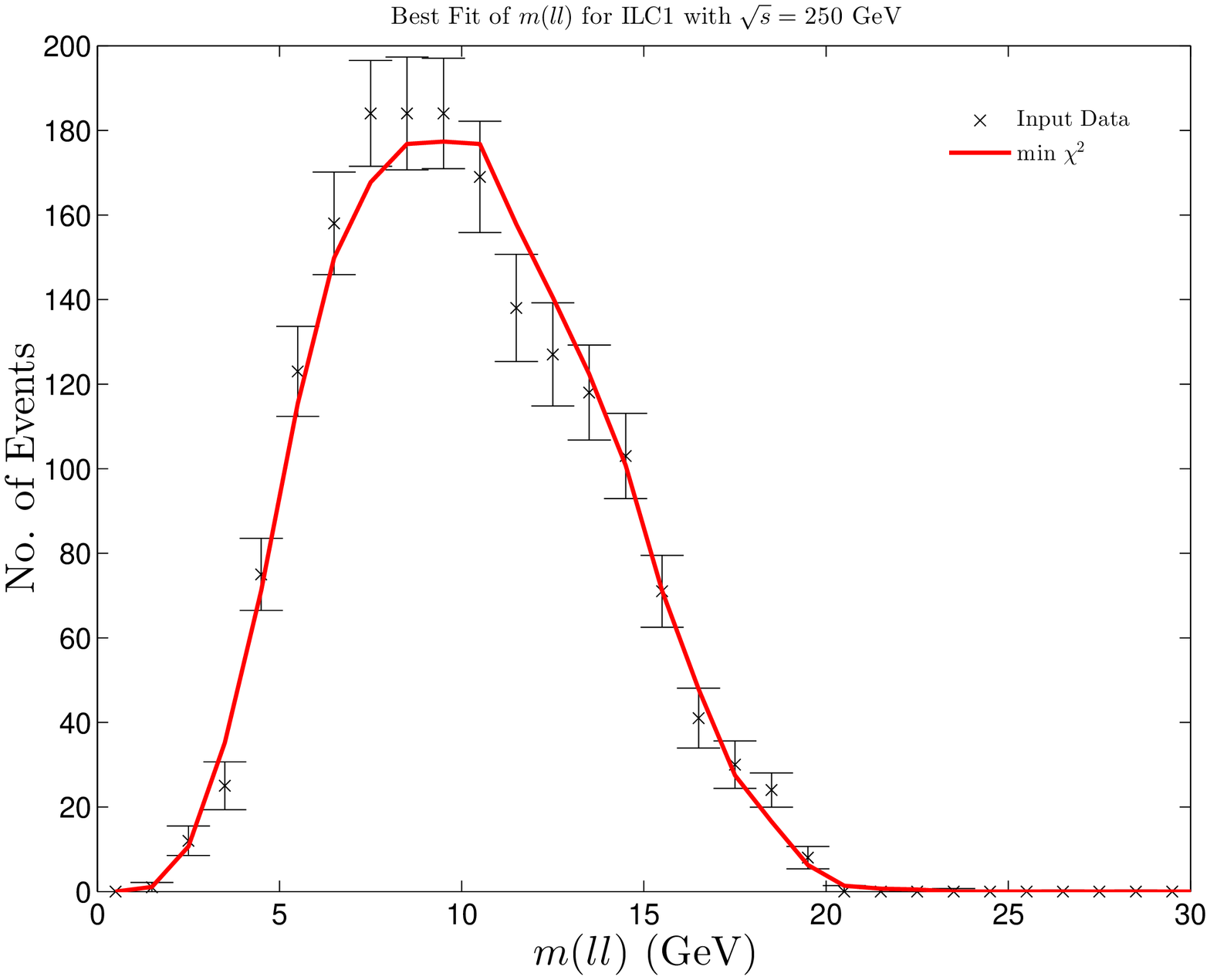}
\caption{In {\it a}), we show values of $\chi^2$ vs. $m(\ell^+\ell^- )$
from 100~fb$^{-1}$ of OS/SF dilepton ILC1 ``data'' from $\tz_1\tz_2$
production fit to the shapes from various ``theory'' templates, as
described in the text.  In {\it b}), we show the ILC1 ``data'' for the
$m(\ell^+\ell^-)$ distribution from $\tz_1\tz_2$ events along with
statistical error for 100~fb$^{-1}$. The solid curve shows the best fit
to these ``data''.  }
\label{fig:chi2_gap}}
%

Once the mass gap is known, it is possible to extract the neutralino
mass value via a fit to the $E(\ell^+\ell^-)$ distribution because the
energy of the daughter leptons (but not their invariant mass) depends 
on the boost of the parent $\tz_2$. 
 We use the same procedure described above to fit the 100
fb$^{-1}$ OS/SF dilepton $E(\ell^+\ell^- )$ using theory templates
with different values of $m_{\tz_2}$, but with
$m_{\tz_2}-m_{\tz_1}$ fixed at 21~GeV.  The corresponding values of 
$\chi^2$ along with the parabolic fit
is shown in Fig.~\ref{fig:chi2ll}{\it a}. We find that
$m_{\tz_2}$ is measured as 
\bi
\item $m_{\tz_2}=123.7\pm 0.2$~GeV ($1\sigma$),
\ei
a 0.2\% measurement.
\FIGURE[tbh]{
\includegraphics[width=12cm,clip]{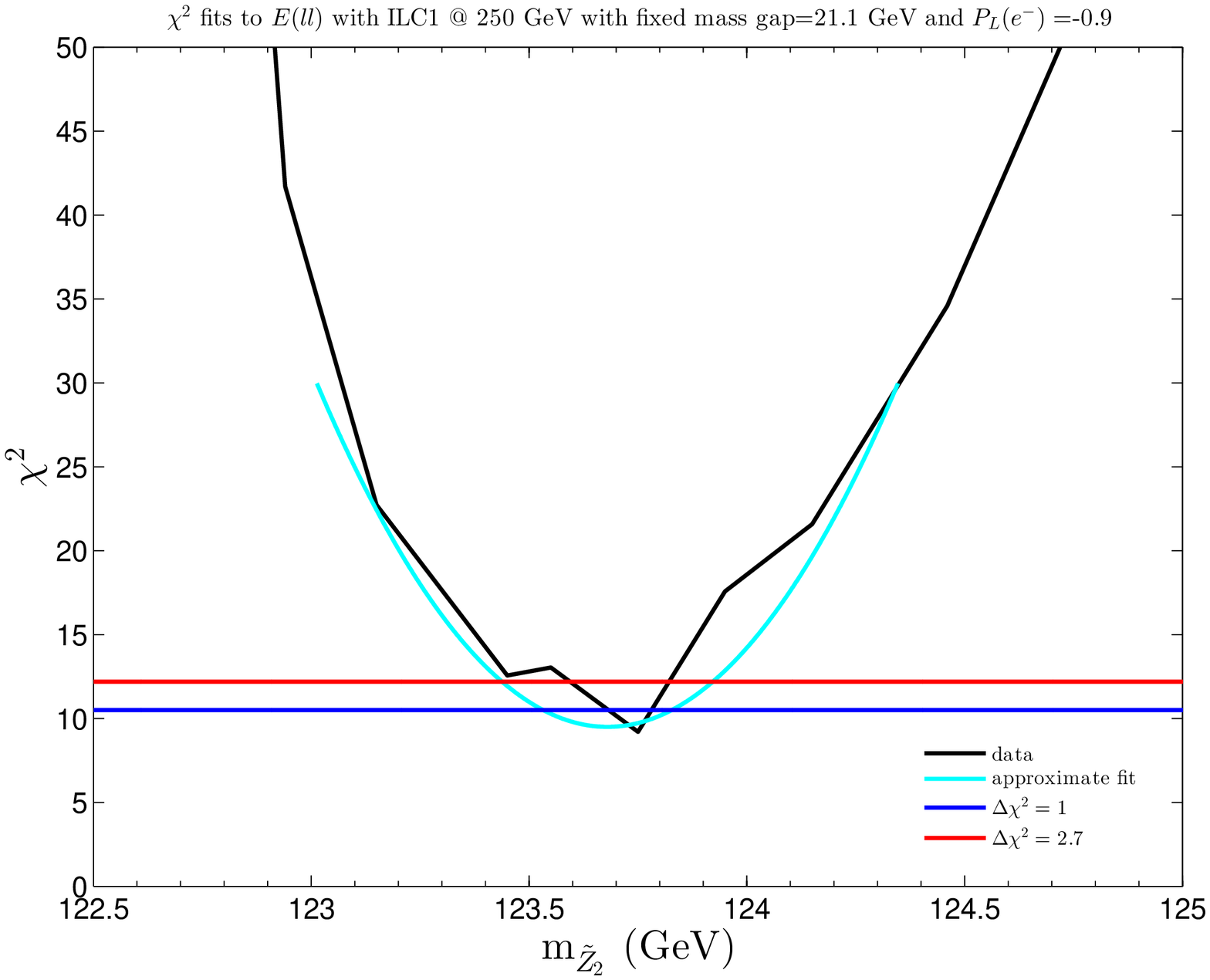}
\includegraphics[width=12cm,clip]{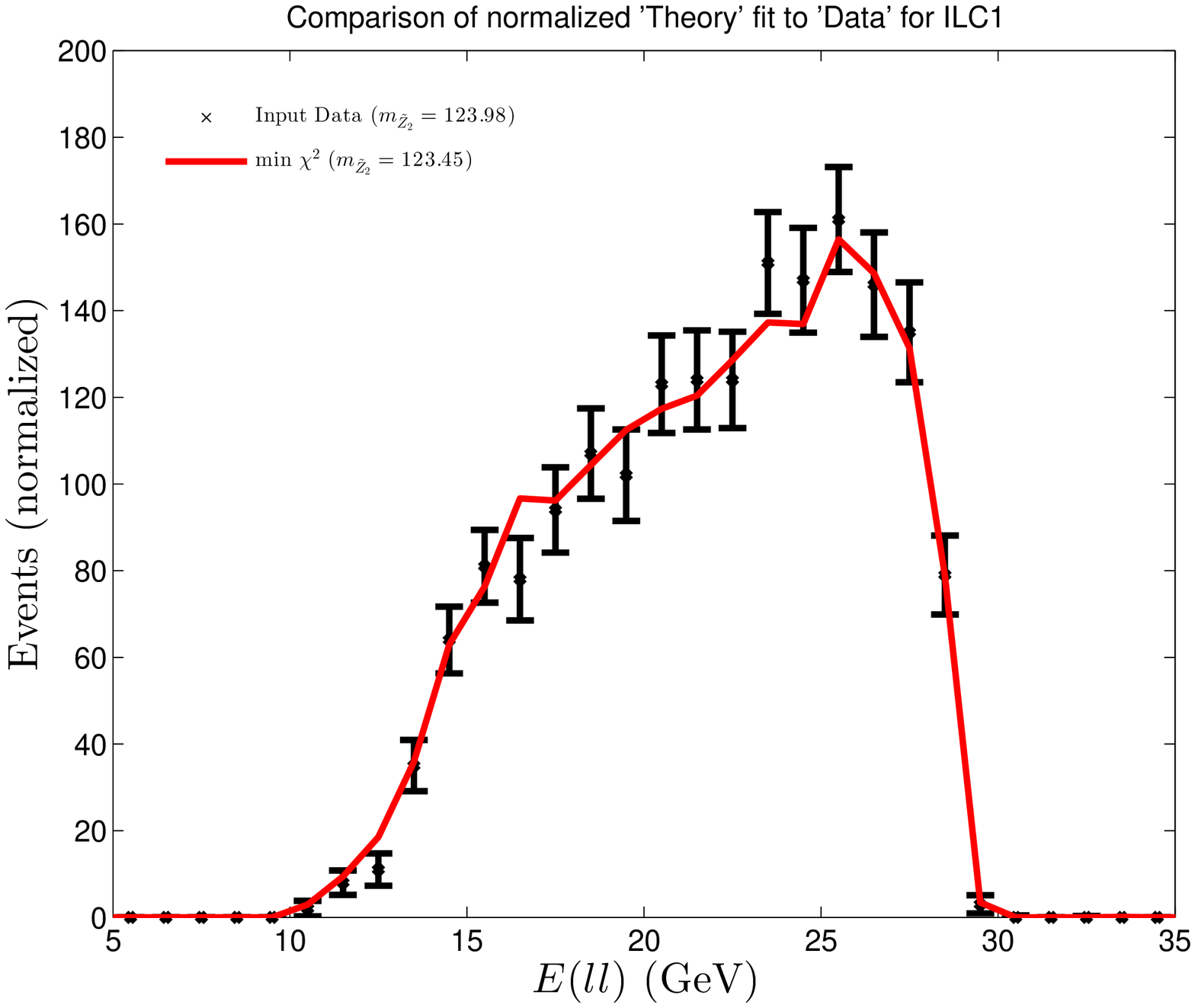}
\caption{In {\it a}), we show fitted values of $\chi^2$ found from
matching 100~fb$^{-1}$ of OS/SF dilepton ``data'' from $\tz_1\tz_2$
production to various ``theory'' distributions generated from varying
$m_{\tz_2}$ while keeping $m_{\tz_2}-m_{\tz_1}$ fixed at 21~GeV.  In
{\it b}), we show the distribution in $E(\ell^+\ell^- )$ from a 100~fb$^{-1}$ 
of OS/SF dilepton ILC1 ``data'' from $\tz_1\tz_2$ production
along with best fit.  }
\label{fig:chi2ll}}
Combining the $m_{\tz_2}-m_{\tz_1}$ and $m_{\tz_2}$ measurements also
gives $m_{\tz_1}$: 
\bi
\item $m_{\tz_1}=102.7\pm 0.3$~GeV (ILC1-dileptons).  
\ei 
This value
serves as a consistency check against the measurement of $m_{\tz_1}$
from chargino pair production, and most importantly, lends support to the
SUSY interpretation of these events. 
The distribution of $E(\ell^+\ell^- )$ data are shown in
Fig.~\ref{fig:chi2ll}{\it b}) along with the corresponding best fit 
depicted by the solid curve.

\subsection{Benchmark ILC2 at $\sqrt{s}=340$~GeV}

Benchmark point ILC2 is more challenging for ILC studies because the
mass gap between $\tw_1/\tz_2$ and the $\tz_1$ is just about 10~GeV,
resulting typically in softer energy release from three-body $\tw_1$ and
$\tz_2$ decays. This mass gap is close to the minimum for RNS models
where $\delew^{-1}> 3$\%.  Of course, since charginos and neutralinos
are heavier, its exploration requires a higher $\sqrt{s}$ to reach
higgsino pair production threshold.  In this case, we perform studies at
$\sqrt{s}=340$~GeV, enough for $\tw_1^+\tw_1^-$ and $\tz_1\tz_2$
production, but just below $t\bar{t}$ threshold. For these higher
$\sqrt{s}$ values, the expected beamstrahlung parameter $\Upsilon$ is
expected to increase to $0.03$, while $\sigma_z$ remains at $0.3$~mm~\cite{tdr3}.

The $E_{vis}$ distribution from signal and background is shown in 
Fig.~\ref{fig:EvisILC2}. Here, we see the ILC2 signal 
restricted to the region with $E_{vis}\alt 30$~GeV, while the
background from  $\gamma\gamma$ collisions is more severe than for
the $\sqrt{s}=250$~GeV case. We impose a cut of
\bi
\item $E_{vis}<30$~GeV.
\ei
\FIGURE[tbh]{
\includegraphics[width=12cm,clip]{etot_340ilc2.eps}
\caption{Distribution in $E_{vis}$ from benchmark ILC2 signal and SM
backgrounds at ILC with $\sqrt{s}=340$~GeV and $\Upsilon =0.03$.  }
\label{fig:EvisILC2}}

Following our earlier analysis, we examine the $\eslt$
in Fig.~\ref{fig:ptm340}.
\FIGURE[tbh]{
\includegraphics[width=12cm,clip]{ptm_340ilc2.eps}
\caption{Distribution in missing transverse energy from $e^+e^-$
collisions at $\sqrt{s}=340$~GeV for signal from the ILC2 benchmark
case, along with SM backgrounds from $e^+e^-$ and $\gamma\gamma$
collisions.  We require $E_{vis}<30$~GeV.  We take beamstrahlung
parameters $\Upsilon =0.03$ and $\sigma_z=0.3$~mm.  }
\label{fig:ptm340}}
We see that, unlike the case of ILC1, the signal never emerges from the
$\gamma\gamma$ background. Clearly additional cuts are necessary for
observability of the signal.

\subsubsection{Chargino pair production for ILC2}
\label{sssec:w1w1ilc2}

To extract a chargino pair production signal from the SM background for benchmark ILC2, we thus require:
\bi
\item $E_{vis}<30$~GeV,
\item $\eslt>10$~GeV,
\item exactly one isolated lepton with $E>5$~GeV and one jet with $E(j)>5$~GeV.  
\ei 
For the case of ILC2, the hadronic energy release from
$\tw_1\to q\bar{q}'\tz_1$ is so small that we almost never produce two
resolvable jets, making chargino mass extraction difficult via continuum
production (although perhaps still possible via threshold scans with sufficient
integrated luminosity). Hence, instead
we focus on the $n(\ell)=1,\ n(jet)=1$ signal. After these requirements,
we plot the transverse plane lepton-jet opening angle which is shown in
Fig.~\ref{fig:thlj}.  Most of the SM background comes from
$\gamma\gamma\to\tau^+\tau^-$ followed by one leptonic and one hadronic
tau decay.  This may be mostly eliminated by requiring 
\bi
\item $\Delta\phi(\ell ,jet)<120^\circ$.
\ei
After this cut we are left with a signal of 7.1~fb whilst SM background
is at the 2.8~fb level, and that all the $\gamma\gamma$ background, which
arises from tau pair production, is eliminated. This should not be
surprising because most taus that decay into 5~GeV jets/leptons will be
significantly boosted, and hence tend to have their decay products nearly
back-to-back in the transverse plane.  We see that a discovery of new
physics might be possible with a data set of just a few fb$^{-1}$ at
ILC340 even in this difficult case.
\FIGURE[tbh]{
\includegraphics[width=12cm,clip]{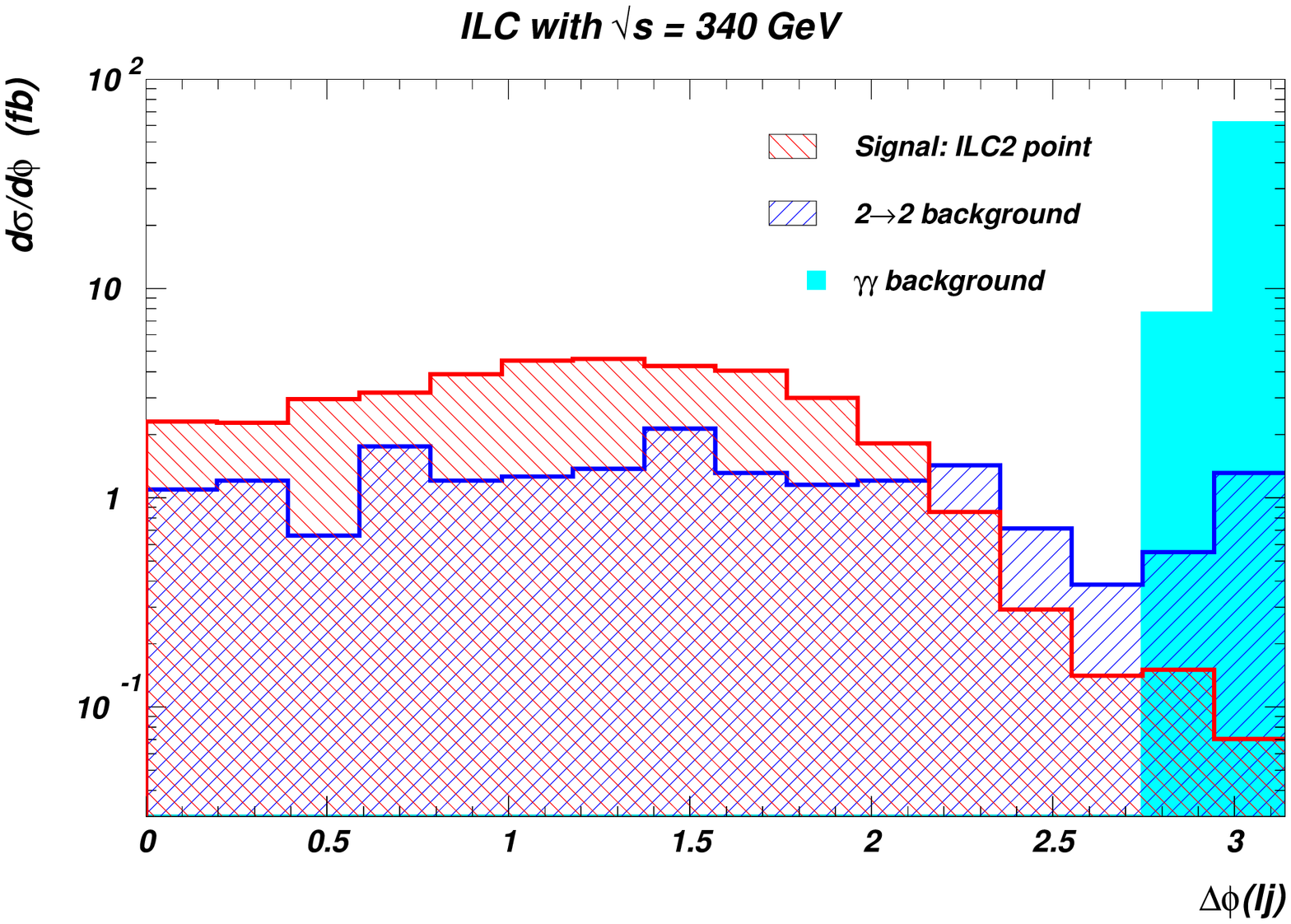}
\caption{Distribution in transverse plane 
opening angle between the isolated lepton and jet 
for the ILC2 signal point at $\sqrt{s}=340$~GeV, and for SM backgrounds.
We require $E_{vis}<30$~GeV and $\eslt>10$~GeV.
We take beamstrahlung parameters $\Upsilon =0.03$ and $\sigma_z=0.3$~mm.
}
\label{fig:thlj}}

\subsubsection{Neutralino pair production for ILC2}
\label{sssec:z1z2ilc2}

As mentioned earlier, if any signal in the $1\ell 1j$ channel just
discussed is to be attributed to higgsino-like charginos of SUSY, we
should also expect a signal from $\tz_1\tz_2$ production as this
reaction  {\em must} have a
comparable production cross section. We are thus led to examine
%
$\tz_1\tz_2$ production for the ILC2 point
with $\sqrt{s}=340$~GeV, where
$\tz_2\to\ell^+\ell^-\tz_1$. This acoplanar dilepton signal may also
allow for neutralino mass reconstruction via continuum production.  
We require 
\bi 
\item $E_{vis}<30$~GeV,
\item a pair of OS/SF leptons, with $n(j)=0$,
\item $\eslt >5$~GeV.
\ei 
For this channel, we operate with mainly right polarized electron beams
with $P_L(e^-)=-0.9$ to reduce backgrounds from $W^+W^-$ and
contamination from $\tw_1^+\tw_1^-$ production.  We next plot the
transverse OS/SF dilepton opening angle in Fig.~\ref{fig:thll}.  To
eliminate the $\gamma\gamma$ background which is once again nearly
back-to-back in the transverse plane, and to improve the
signal-to-background ratio,  we also require: 
\bi
\item $\Delta\phi(\ell^+\ell^-)<90^\circ$.
\ei 
At this point, we have a
signal sample of $2.6$~fb while SM background is at the $0.15$~fb level
with no $\gamma\gamma$ background. Once again, discovery of new physics
is possible with just a few~fb$^{-1}$ of integrated luminosity. 
%
\FIGURE[tbh]{
\includegraphics[width=12cm,clip]{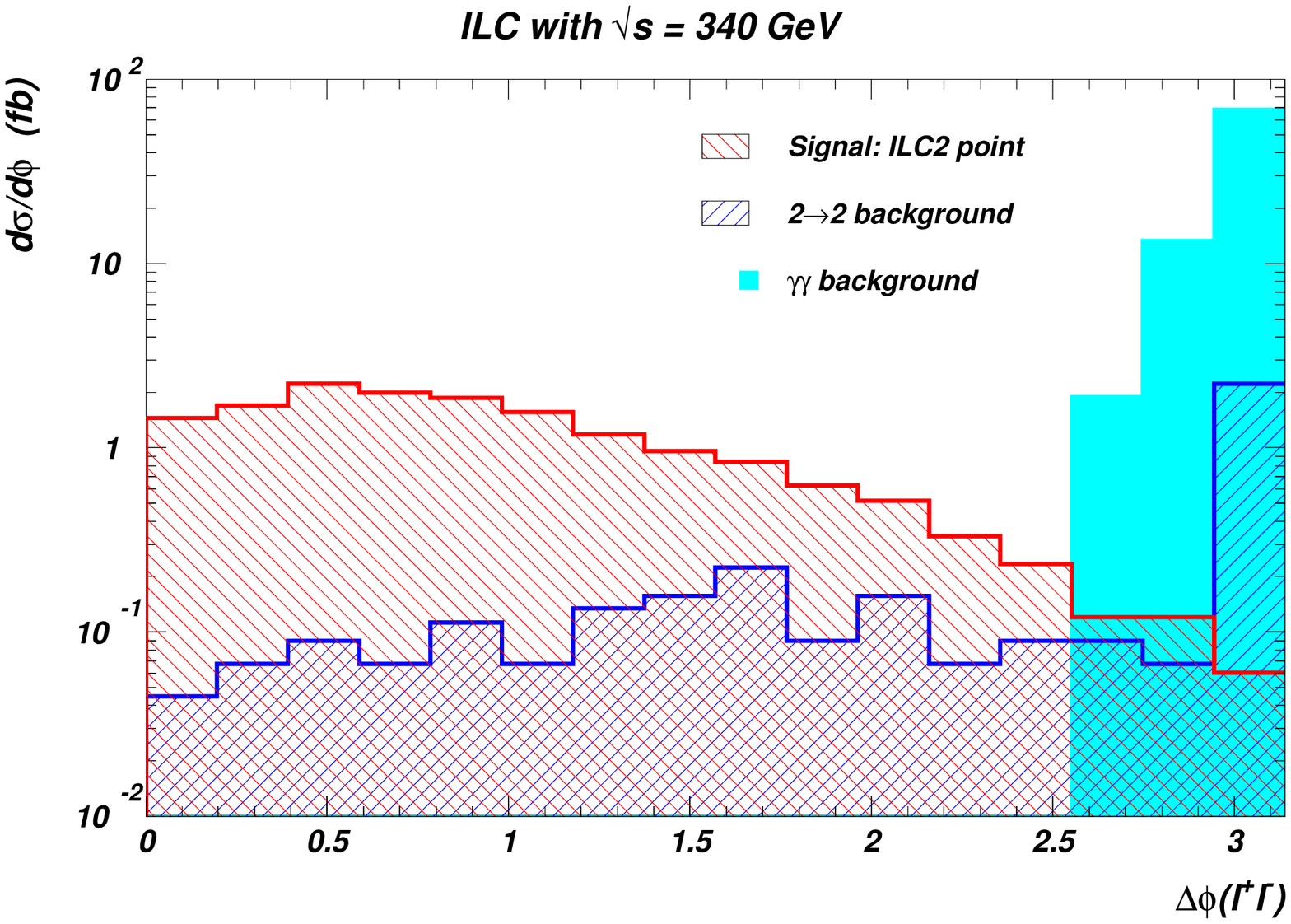}
\caption{Distribution in transverse opening angle between isolated OS/SF
leptons for ILC2 signal at $\sqrt{s}=340$~GeV and for SM backgrounds.
We require $E_{vis}<30$~GeV and $\eslt> 5$~GeV.  We take beamstrahlung
parameters $\Upsilon =0.03$ and $\sigma_z=0.3$~mm.  }
\label{fig:thll}}

For benchmark ILC2, we use  the same procedure that we used for the ILC1 
case study to extract the neutralino masses. We first
fit the normalized theory templates (generated with 1000~fb$^{-1}$ each)
with varying $m_{\tz_2}-m_{\tz_1}$ mass gaps to a 100~fb$^{-1}$ ``data''
distribution.  The various $\chi^2$ values along with a parabolic fit
are shown in Fig.~\ref{fig:chi2_gap_ilc2}{\it a}. As before, we 
have checked that the very large $\chi^2$ values for a mass gap
near the bottom of the parabola arise from extreme values of $m_{\tz_2}$
in the templates. For the ILC2 case, we
find the $m_{\tz_2}-m_{\tz_1}$ mass gap is measured to be 
\bi
\item $m_{\tz_2}-m_{\tz_1} = 9.7\pm 0.2$~GeV ($1\sigma$) 
\ei 
a 2\% measurement.  The data along with best theory fit are shown in
Fig.~\ref{fig:chi2_gap_ilc2}{\it b}.
\FIGURE[tbh]{
\includegraphics[width=12cm,clip]{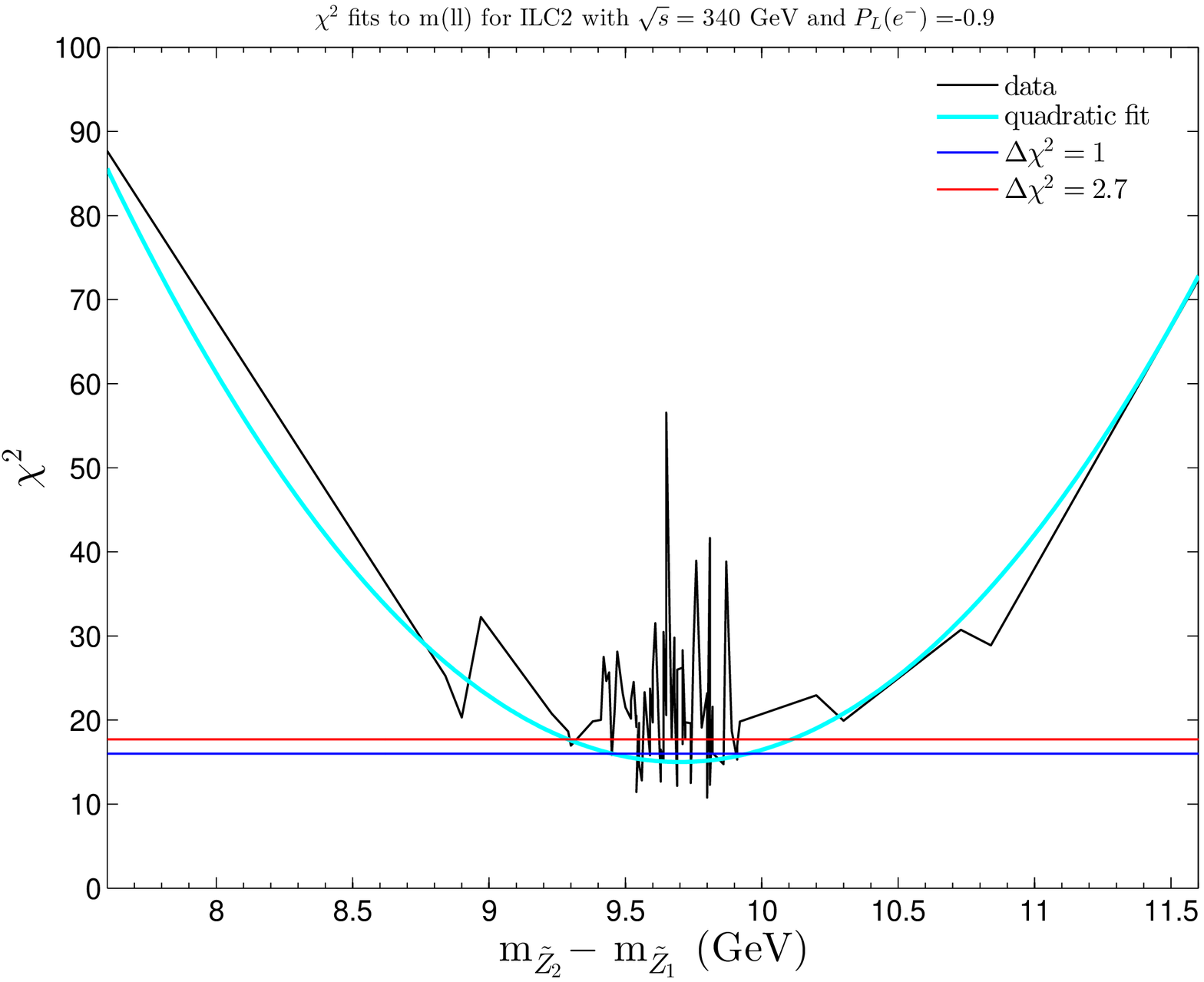}
\includegraphics[width=12cm,clip]{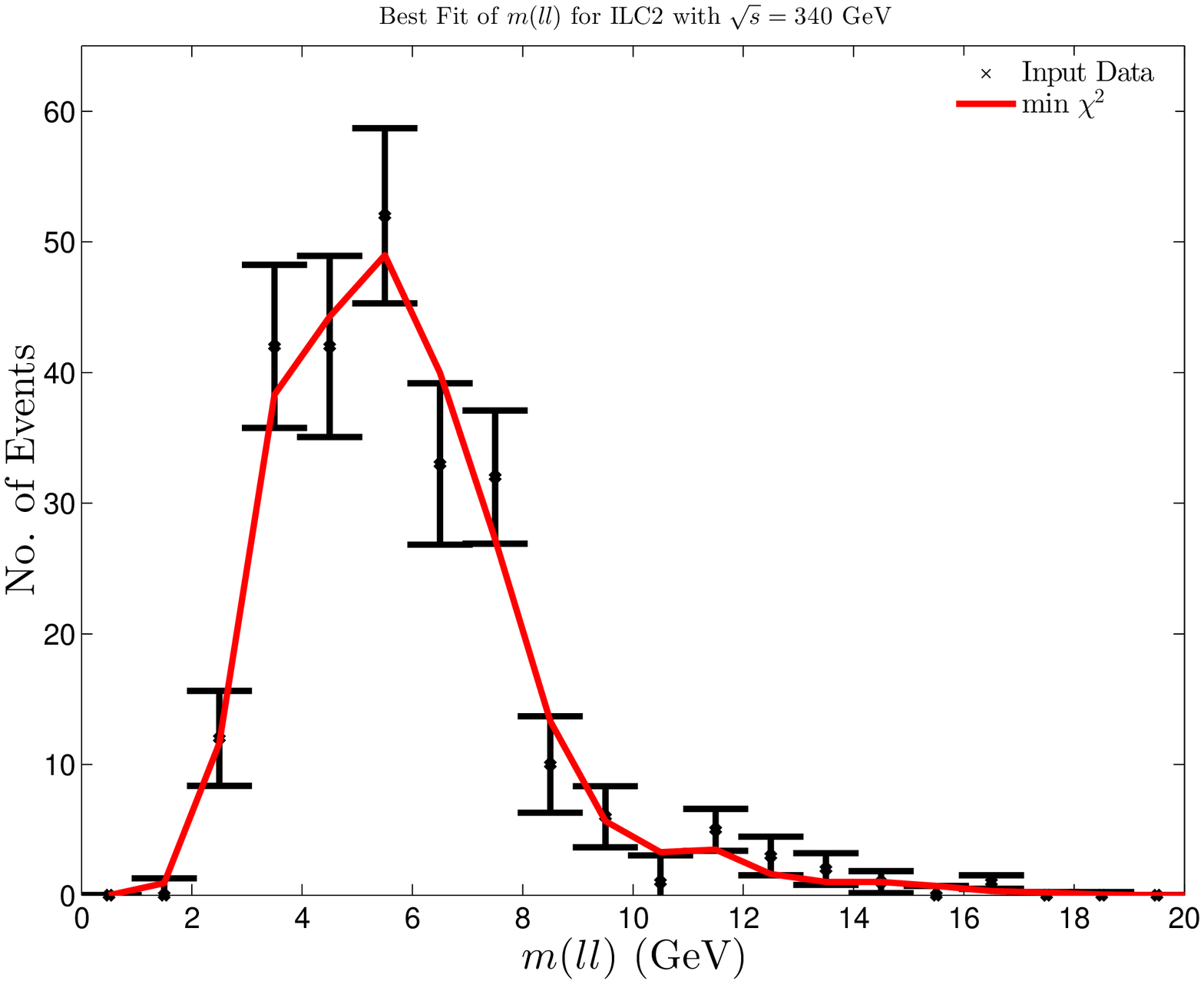}
\caption{In {\it a}), we show $\chi^2$ values vs. $m(\ell^+\ell^- )$
from 100~fb$^{-1}$ of OS/SF dilepton ILC2 ``data'' from $\tz_1\tz_2$
production fit to theory along with a best-fit parabola.  In {\it b}),
we show the distribution in $m(\ell^+\ell^- )$ from a 100~fb$^{-1}$ of
OS/SF dilepton ILC2 ``data'' from $\tz_1\tz_2$ production along with
best fit.  }
\label{fig:chi2_gap_ilc2}}

Next,  keeping the mass gap fixed near 9.7~GeV, we generate theory
templates for the $E(\ell^+\ell^- )$ distributions
from $\tz_1\tz_2$ production with 10 times the statistics of
``data'' but with varying $m_{\tz_2}$ values and  fit the shapes of
these to the corresponding ``data'' distribution as before. 
In Fig.~\ref{fig:chi2ll_ilc2}, we show the $\chi^2$ values along with the  parabolic fit. 
We find a measurement of 
\bi
\item $m_{\tz_2}=158.5\pm 0.4$~GeV ($1\sigma$).
\ei
The $E(\ell^+\ell^-)$ distribution for ``data'' along with best fit
theory are shown in Fig.~\ref{fig:chi2ll_ilc2}{\it b}.  
\FIGURE[tbh]{
\includegraphics[width=12cm,clip]{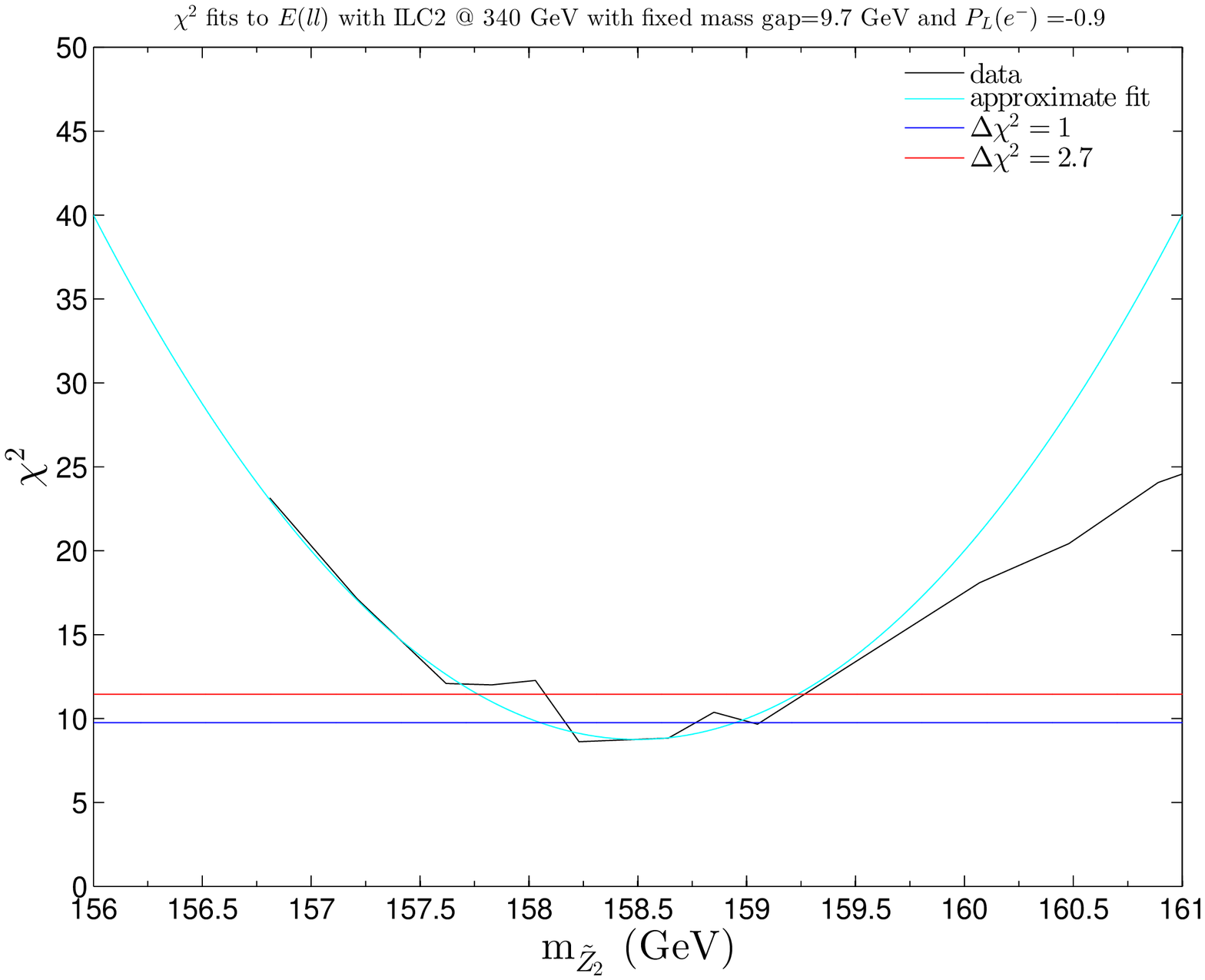}
\includegraphics[width=12cm,clip]{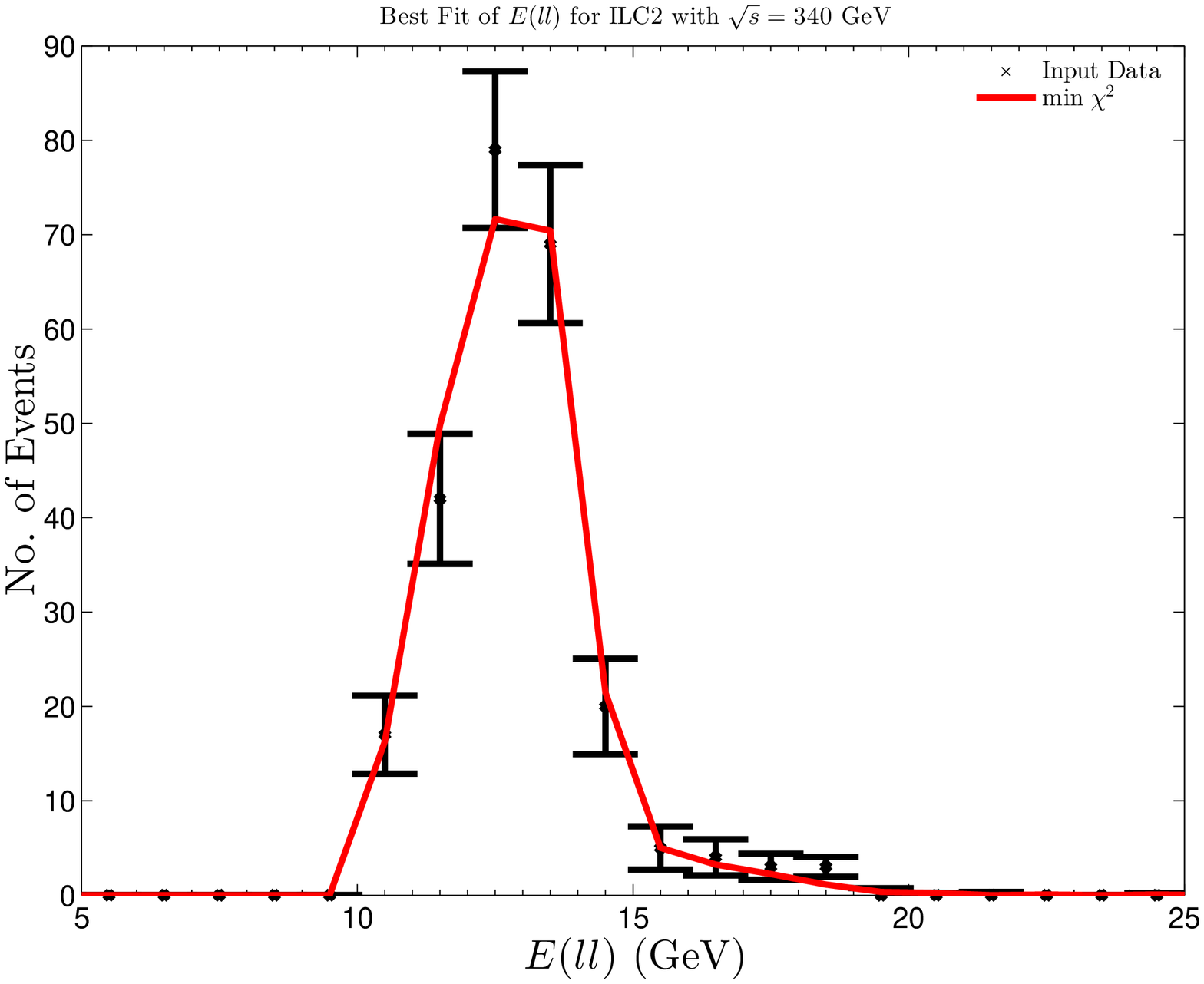}
\caption{In {\it a}), we show values of $\chi^2$ found from matching 100~fb$^{-1}$ 
of OS/SF dilepton ILC2 ``data'' from $\tz_1\tz_2$ production
to various ``theory'' distributions generated from varying $m_{\tz_2}$
while keeping $m_{\tz_2}-m_{\tz_1}$ fixed at 9.7~GeV.  In {\it b}), we
show the distribution in $E(\ell^+\ell^- )$ from a 100~fb$^{-1}$ of
OS/SF dilepton ILC2 ``data'' from $\tz_1\tz_2$ production along with
best fit.  }
\label{fig:chi2ll_ilc2}}
By combining the $m_{\tz_2}$ and mass gap measurements, we find
\bi
\item $m_{\tz_1}=148.8\pm 0.5$~GeV (ILC2-dileptons).
\ei

\section{Conclusions}
\label{sec:conclude}

Naturalness arguments imply small $|\mu|$, and concomitantly four light
higgsino-like states with masses $\sim 100-300$~GeV, the closer to $M_Z$
the more natural. Because of the small energy release in their decays,
direct higgsino production may be difficult-to-impossible to detect at
LHC14, while the International Linear $e^+e^-$ Collider would be a
higgsino factory and serve as a SUSY discovery machine so long as
$\sqrt{s}>2m(higgsino)$.  We investigated two benchmark scenarios:
ILC1 with lighter higgsinos $\sim 120$~GeV and mass gap $\sim 15-22$~GeV
relative to the LSP, and ILC2 with heavier higgsinos $\sim 150$~GeV but
with a mass gap of just 10~GeV, close to the minimum possible in  models
with no worse than 3\% fine-tuning.

For both these cases, the chargino pair and neutralino pair signals
should be seen above usual SM $2\to 2$ background and $\gamma\gamma$
induced events via a combination of specially devised $E_{vis}$,
$\eslt$, angle and topology cuts with an integrated luminosity of just a
few fb$^{-1}$.  The signal will be characterized by low $E_{vis}$
($\lesssim 30-50$~GeV)  plus $\eslt$ events indicating the production
of heavy parents that decay into an invisible partner with a mass just
10-20~GeV lighter.  Observation of a signal in both $jet(s) + \ell$ and OS/SF
dilepton channels at the expected rates will point to the production of
higgsinos that are the hallmark of natural SUSY models.  For ILC1, the
$\ell+jets$ signal allows for a continuum measurement of $m_{\tw_1}$ and
$m_{\tz_1}$ to $\sim 2\%$ accuracy assuming a canonical value of
100~fb$^{-1}$ of integrated luminosity.  
The neutralino pair production reaction
can be seen above the background in the OS/SF, same-side dilepton signal
which allows for the mass gap $m_{\tz_2}-m_{\tz_1}$ to be measured via
the $m(\ell^+\ell^-)$ distribution to $\sim 1\%$ accuracy while
$m_{\tz_2}$ can be measured to sub-GeV precision.

The more challenging ILC2 point allows for chargino pairs to be seen
above background, but the smaller $m_{\tw_1}-m_{\tz_1}$ mass gap
makes dijet resolution very difficult so that a continuum mass
measurement via $E(jj)$ is not possible using our simple
methods. The OS/SF same-side dilepton measurement still remains viable in the case of
ILC2 where the mass gap $m_{\tz_2}-m_{\tz_1}$ can be measured to $\sim
2\%$ accuracy and $m_{\tz_2}$ can be again measured to sub-GeV precision.

Although we have performed our analysis using the RNS model as a guide,
our results should be applicable to all models with light higgsinos. We
are encouraged that even the difficult point with the smallest mass gap
for 150~GeV higgsinos allows not only detection, but also precision mass
measurements, even at a centre-of-mass energy just modestly above the
production threshold. This leads us to infer that an electron-positron
collider will serve as a definitive probe of the idea of naturalness in
all SUSY models where the superpotential $\mu$-term is the dominant
contribution to higgsino masses. In particular, from the dashed contour
in Fig.~\ref{fig:gap}, we conclude that ILC600 will either discover or
decisively exclude models where fine-tuning is worse than 3\%.
Precision measurements that can be made
at ILC will definitively show that higgsino pair production is indeed
occuring, and will allow us to measure with high precision at least the
higgsino mass scale and associated mass gaps. Such a discovery would
not only confirm SUSY but also strongly indicate that $W$, $Z$ and $h$ mass
scales arise in a natural way via a link to nearby light higgsinos.


\acknowledgments

We thank Peisi Huang for collaboration in the early stages of this work
and we thank T. Barklow, N. Graf and J. List for discussions.
This work was supported in part by the U.S. Department of Energy.


%


\begin{thebibliography}{99}
%
\bibitem{atlas_h} G.~Aad {\it et al.}  [ATLAS Collaboration], 
 \plb{716}{2012}{1}.
%
\bibitem{cms_h} S.~Chatrchyan {\it et al.}  [CMS Collaboration],  
\plb{716}{2012}{30}.
%
\bibitem{atlas_susy}
  G.~Aad {\it et al.}  [ATLAS Collaboration],
  \arXivid{1208.0949}~[hep-ex];
  ATLAS-CONF-2013-047; 
  ATLAS-CONF-2013-062.
%
\bibitem{cms_susy}
  S.~Chatrchyan {\it et al.}  [CMS Collaboration],
  \arXivid{1207.1798}~[hep-ex];  
  \arXivid{1402.4770}~[hep-ex]; 
  \epjc{73}{2013}{2568} [\arXivid{1303.2985}~[hep-ex]].
%
\bibitem{Shifman:2012na}
  M.~Shifman,
  Mod.\ Phys.\ Lett.\ A {\bf 27} (2012) 1230043.
%
\bibitem{craig} N.~Craig,
  \arXivid{1309.0528}~[hep-ph].
%
\bibitem{ltr} H.~Baer, V.~Barger, P.~Huang, A.~Mustafayev and X.~Tata,
  \prl{109}{2012}{161802}.
%
\bibitem{DEWsug} H.~Baer, V.~Barger, P.~Huang, D.~Mickelson, A.~Mustafayev and X.~Tata,
  \prd{87}{2013}{035017}.
%
\bibitem{rns} H.~Baer, V.~Barger, P.~Huang, D.~Mickelson, A.~Mustafayev and X.~Tata,
  \prd{87}{2013}{115028}
  [\arXivid{1212.2655}~[hep-ph]].
%
\bibitem{snowmass1} H.~Baer, V.~Barger, P.~Huang, D.~Mickelson, A.~Mustafayev and X.~Tata,
  \arXivid{1306.2926}~[hep-ph]. 
%
\bibitem{comp} H.~Baer, V.~Barger and D.~Mickelson,
  \prd{88}{2013}{095013}.
%
\bibitem{am_xt} A.~Mustafayev and X.~Tata,
  \arXivid{1404.1386}~[hep-ph].
%
\bibitem{dew} H.~Baer, V.~Barger, D.~Mickelson and M.~Padeffke-Kirkland,
  \arXivid{1404.2277}~[hep-ph].
%
\bibitem{ccn} K.~L.~Chan, U.~Chattopadhyay and P.~Nath,
  \prd{58}{1998}{096004} 
  [hep-ph/9710473];
S.~Akula, M.~Liu, P.~Nath and G.~Peim,
  \plb{709}{2012}{192};
M.~Liu and P.~Nath,
  \prd{87}{2013}{095012}. 
%
\bibitem{fp} J.~L.~Feng, K.~T.~Matchev and T.~Moroi,
  \prd{61}{2000}{075005}; 
J.~L.~Feng and K.~T.~Matchev,
  \prd{63}{2001}{095003};
J.~L.~Feng, K.~T.~Matchev and D.~Sanford,
  \prd{85}{2012}{075007};
J.~L.~Feng and D.~Sanford,
  \prd{86}{2012}{055015}.
%
\bibitem{golden}
  M.~Perelstein and C.~Spethmann,
  \jhep{0704}{2007}{070}.
%
\bibitem{perel} M.~Perelstein and B.~Shakya,
  \jhep{1110}{2011}{142}; 
 M.~Perelstein and B.~Shakya,
  \prd{88}{2013}{075003}.
%
\bibitem{martin}  S.~P.~Martin,
  \prd{75}{2007}{115005};
J.~E.~Younkin and S.~P.~Martin,
  \prd{85}{2012}{055028};
S.~P.~Martin,
   \prd{89}{2014}{035011}.
%
\bibitem{nevzorov} S.~F.~King, M.~Muhlleitner and R.~Nevzorov,
  \npb{860}{2012}{207}.
%
\bibitem{guido} G.~Altarelli,
  Phys.\ Scripta T {\bf 158} (2013).
%
\bibitem{brust} C.~Brust, A.~Katz, S.~Lawrence and R.~Sundrum,
  \jhep{1203}{2012}{103}.
%
\bibitem{deg} H.~Baer, V.~Barger, M.~Padeffke-Kirkland and X.~Tata,
  \prd{89}{2014}{037701}.
%
\bibitem{dine}  M.~Dine, A.~Kagan and S.~Samuel, \plb{243}{1990}{250};
A.~Cohen, D.~B.~Kaplan and A.~Nelson, \plb{388}{1996}{588};
N.~Arkani-Hamed and H.~Murayama, \prd{56}{1997}{R6733};
T.~Moroi and M.~Nagai,
  \plb{723}{2013}{107}.
%
\bibitem{lhc}  H.~Baer, V.~Barger, P.~Huang, D.~Mickelson, A.~Mustafayev, W.~Sreethawong and X.~Tata,
  \jhep{1312}{2013}{013}.
%
\bibitem{SSdB}
H.~Baer, V.~Barger, P.~Huang, D.~Mickelson, A.~Mustafayev, W.~Sreethawong and X.~Tata,
  \prl{110}{2013}{151801}.
%
\bibitem{chan} C.~Han, A.~Kobakhidze, N.~Liu, A.~Saavedra, L.~Wu and J.~M.~Yang,
  \jhep{1402}{2014}{049}.
%
\bibitem{mono} H.~Baer, A.~Mustafayev and X.~Tata,
  \prd{89}{2014}{055007}.
%
\bibitem{sz} P.~Schwaller and J.~Zurita, \jhep{1403}{2014}{060}.
%
\bibitem{kribs} Z.~Han, G.~Kribs, A.~Martin and A.~Menon,
  \arXivid{1401.1235}~[hep-ph].
%
\bibitem{tdr}
  H.~Baer, T.~Barklow, K.~Fujii, Y.~Gao, A.~Hoang, S.~Kanemura, J.~List and H.~E.~Logan {\it et al.},
  \arXivid{1306.6352}~[hep-ph].
%
\bibitem{girardello} L.~Girardello and M.~T.~Grisaru, \npb{194}{1982}{65};
  I.~Jack and D.~R.~T.~Jones, \plb{457}{1999}{101}.
%
\bibitem{snowmass} H.~Baer, M.~Berggren, J.~List, M.~M.~Nojiri,
  M.~Perelstein, A.~Pierce, W.~Porod and T.~Tanabe, 
  \arXivid{1307.5248}~[hep-ph].
%
\bibitem{bbh} H.~Baer, V.~Barger and P.~Huang,
  JHEP {\bf 1111} (2011) 031. 
%
\bibitem{jlc}
  T.~Tsukamoto, K.~Fujii, H.~Murayama, M.~Yamaguchi and Y.~Okada,
  \prd{51}{1995}{3153}.
%
\bibitem{bmt} H.~Baer, R.~Munroe and X.~Tata,
  \prd{54}{1996}{6735} [Erratum-ibid.\ D {\bf 56} (1997) 4424]. 
%
\bibitem{tadas1} H.~Baer, A.~Belyaev, T.~Krupovnickas and X.~Tata,
  \jhep{0402}{2004}{007}.

\bibitem{tadas2} H.~Baer, T.~Krupovnickas and X.~Tata,
  \jhep{0406}{2004}{061}.
%
\bibitem{Han:2013kza}
  T.~Han, S.~Padhi and S.~Su,
  \prd{88}{2013}{115010};
M.~Berggren, T.~Han, J.~List, S.~Padhi, S.~Su and T.~Tanabe,
  \arXivid{1309.7342}~[hep-ph].
%
\bibitem{Berggren:2013vfa}
  M.~Berggren, F.~Brümmer, J.~List, G.~Moortgat-Pick, T.~Robens, K.~Rolbiecki and H.~Sert,
  \epjc{73}{2013}{2660}.
%
\bibitem{ilcbm} H.~Baer and J.~List,
  \prd{88}{2013}{055004}.
%
\bibitem{isasugra} H.~Baer, C.~H.~Chen, R.~Munroe, F.~Paige and X.~Tata, \prd{51}{1995}{1046}; 
H.~Baer, J.~Ferrandis, S.~Kraml and W.~Porod, \prd{73}{2006}{015010}.
%
\bibitem{isajet} ISAJET, by H.~Baer, F.~Paige, S.~Protopopescu and
X.~Tata, \hepph{0312045}.
%
\bibitem{nuhm2}  D.~Matalliotakis and H. P.~Nilles, \npb{435}{1995}{115}; 
V.~Berezinsky {\it et al.} {\em Astropart. Phys.}  {\bf 5} (1996) 1; 
P.~Nath and R.~Arnowitt, \prd{56}{1997}{2820}; 
J.~Ellis, K.~Olive and Y.~Santoso, \plb{539}{2002}{107};
J.~Ellis, T.~Falk, K.~Olive and Y.~Santoso, \npb{652}{2003}{259};
H.~Baer, A.~Mustafayev, S.~Profumo, A.~Belyaev and
  X.~Tata, \prd{71}{2005}{095008} and \jhep {0507}{2005}{065}, 
and references therein.
%
\bibitem{bbm} H.~Baer, V.~Barger and D.~Mickelson,
  \plb{726}{2013}{330}.
%
\bibitem{dfsz} K.~J.~Bae, H.~Baer and E.~J.~Chun,
  Phys.\ Rev.\ D {\bf 89} (2014) 031701 and JCAP {\bf 1312} (2013) 028.
%
\bibitem{bbkmt} H.~Baer, A.~Bartl, D.~Karatas, W.~Majerotto and X.~Tata,
  \ijmpa{4}{1989}{4111}.
%
\bibitem{loop} H.~Komatsu and H.~Kubo, \plb{157}{1985}{90} and \npb
  {263}{1986}{265}; H.~Haber and D.~Wyler, \npb {323}{1989}{267};
  H.~Baer and T.~Krupovnickas, \jhep{0209}{2002}{038}.
%
\bibitem{brem}E.~A.~Kuraev and V.S~Fadin, {\em Sov. J. Nucl. Phys. }
  {\bf 41} (1985) 466.
%
\bibitem{beam} P.~Chen, \prd{46}{1992}{1186}; M.~Peskin, SLAC-TN-04-032.
%
\bibitem{tdr3} 
  C.~Adolphsen, M.~Barone, B.~Barish, K.~Buesser, P.~Burrows, J.~Carwardine, J.~Clark and Hélèn.~M.~Durand {\it et al.},
  \arXivid{1306.6328}~[physics.acc-ph].
%
\bibitem{drees} M.~Drees and R.~M.~Godbole, 
  \prd{50}{1994}{3124}.
%
\bibitem{wss} See, {\it e.g.} {\em  Weak Scale Supersymmetry}, H.~Baer
  and X.~Tata (Cambridge, 2006), where all relevant couplings may be found. 
%
\bibitem{threshold}
  G.~A.~Blair,
  eConf C {\bf 010630} (2001) E3019; 
  for corresponding studies at a muon collider see 
V.~D.~Barger, M.~S.~Berger and T.~Han,
  \prd{59}{1999}{71701}.
%
\bibitem{kadala} R.~Kadala, Ph. D. dissertation, \arXivid{1205.1267}.
%
\bibitem{kn} R.~Kitano and Y.~Nomura, \prd{73}{2006}{095004}.
%

\end{thebibliography}
\end{document}